\documentclass[prb,amssymb,twocolumn,superscriptaddress,showpacs]{revtex4}
\usepackage{bm}
\usepackage{graphicx}
\usepackage{color}
\usepackage{amsmath}

\begin{document}

\title{
Microscopics of disordered two-dimensional electron gases under high magnetic
fields: Equilibrium properties and dissipation in the hydrodynamic regime}

\author{Thierry Champel}
\affiliation{Universit\'{e} Joseph Fourier, Laboratoire de Physique et
Mod\'{e}lisation des Milieux Condens\'{e}s, CNRS, B.P. 166,
25 Avenue des Martyrs, 38042 Grenoble Cedex 9, France}

\author{Serge Florens}
\affiliation{Institut N\'{e}el, CNRS and Universit\'{e} Joseph Fourier,
B.P. 166, 25 Avenue des Martyrs, 38042 Grenoble Cedex 9, France }

\author{L\'{e}onie Canet}
\affiliation{Universit\'{e} Joseph Fourier, Laboratoire de Physique et
Mod\'{e}lisation des Milieux Condens\'{e}s, CNRS, B.P. 166,
25 Avenue des Martyrs, 38042 Grenoble Cedex 9, France}

\date{\today}

\begin{abstract}
We develop in detail a new formalism [as a sequel to the work of T. Champel and S. Florens, Phys. Rev. B {\bf 75}, 245326
(2007)] that is well-suited for treating quantum problems involving
slowly varying potentials at high magnetic fields in two-dimensional electron
gases. For an arbitrary smooth potential we show that the electronic Green's function
is fully determined by closed recursive expressions that take the form of a high
magnetic field expansion in powers of the magnetic length $l_{B}$.
For illustration we determine entirely Green's function at order $l_{B}^3$,
which is then used to obtain quantum expressions for the local charge and current
electronic densities at equilibrium. Such results are valid at high but finite magnetic
fields and for arbitrary temperatures, as they take into account Landau level
mixing processes and wave function broadening.
We also check the accuracy of our general functionals against the exact solution
of a one-dimensional parabolic confining potential, demonstrating the controlled
character of the theory to get equilibrium properties.
Finally, we show that transport in high magnetic fields can be described
hydrodynamically by a local equilibrium regime and that dissipation mechanisms
and quantum tunneling processes are intrinsically included at the
microscopic level in our high magnetic field theory. We calculate microscopic expressions for the local conductivity tensor, which possesses both transverse and longitudinal components,
providing a microscopic basis for the understanding of dissipative features in
quantum Hall systems.
\end{abstract}


\maketitle

\section{Introduction}

\subsection{General motivation}

Almost thirty years after the discovery of the quantum Hall
effect~\cite{vonK,vonKlitzing2005}, two-dimensional electron gases under magnetic fields continue to attract a considerable interest both experimentally and
theoretically, and have revealed a rich world of surprising physics. 
Newly
discovered features concern, e.g., the zero resistance states under microwave 
illumination~\cite{Mani} and the peculiar Landau-level quantization in 
graphene~\cite{Novoselov}. Concerning the integer quantum Hall effect (IQHE) itself,
 direct local imaging
techniques~\cite{Weitz,Ilani,Steele,Weis2007} have revealed new electron-electron correlation phenomena and  allowed a finer understanding of the microscopic ingredients at work.

On the theoretical side, the quantization of the Hall resistance observed in the IQHE relies on the 
understanding of the quantum motion of charged particles in a two-dimensional 
disordered electrostatic landscape in the presence of a strong perpendicular 
magnetic field
\cite{Kazarinov1982,Trugman1983,Apenko1983,Joynt,Apenko1984,Apenko1985,Shapiro1986,Prange1987,Chalker1988,Huckestein1995}. 
As the main effects of the electron-electron interaction can be taken into
account in the integer regime at the single-particle level, using a Hartree
approach to screening \cite{Chklovskii1992,Lier1994}, the calculation of equilibrium
properties, such as the local electronic density and the distribution of
permanent currents throughout the sample, can be carried out from a one-particle
random Schr\"{o}dinger equation. The precise resolution of this problem
constitutes the first and important step toward a microscopic description that 
underlies the more complex nonequilibrium phenomena of the quantum Hall effect 
in its generality.

Despite the overall good understanding gained after several decades of research
\cite{Trugman1983,Joynt,Apenko1984,Apenko1985,Shapiro1986,Prange1987,Chalker1988,Huckestein1995,Chklovskii1992,Lier1994,
Laughlin1981,Halperin1982,MacDonald1984,Buttiker1988,Streda1982},
a simple and general microscopic approach for the physics of quantum Hall systems
is surprisingly still lacking.
Computer-based simulations have been proposed \cite{Wexler1994,Sohrmann2007},
but rely on heavy numerics in the case of two-dimensional disordered potentials,
and are not well suited for the computation of out-of-equilibrium properties.
Even in the linear response regime, they are certainly unable to address minute
aspects such as the tiny deviations to the Hall resistance quantization.
Analytical approaches are better adapted to formulate out-of-equilibrium calculations,
but face the need to reliably handle in a self-consistent screening theory both
the formation of Landau energy levels at quantizing magnetic fields
and the complexity of the random potential.

At present, the theory of the integer quantum Hall effect relies on two main
cornerstones, which do await a unified description.
On the one hand, equilibrium density profiles are generally computed within a
semiclassical Thomas-Fermi approach \cite{Chklovskii1992,Lier1994,Cooper1993,Guven2003},
leading to a description of the quantum Hall liquid in terms of compressible and
incompressible regions. These calculations have however proved to be only
qualitatively accurate \cite{Siddiki2004}, and fail at low temperature, where
quantum broadening due to the electronic wave function becomes important.
Transport properties are on the other hand conveniently formulated in the B\"{u}ttiker
edge state formalism \cite{Buttiker1988,Chklovskii1992,Cooper1993}, which nevertheless needs input from
more microscopic calculations of the bulk properties. This scattering theory
becomes also very cumbersome to describe dissipative features of macroscopic
samples. An alternative successful semiclassical approach to transport \cite{Ruzin1993,Dykhne1994,Simon1994,Guven2003,Siddiki2004} assumes local Ohm's law at a phenomenological level. 
A simple and controlled quantum approach to both equilibrium and out-of-equilibrium
properties of quantum Hall fluids is thus clearly needed.

\subsection{Review of the high magnetic field approaches: The semiclassical limit}
A popular approach to the IQHE is the high magnetic field
limit, in which case the center-of-mass motion of the electron becomes
essentially classical. A quantum description is only kept for the orbital
effects associated with the Landau-level formation
\cite{Trugman1983,Joynt,Apenko1984,Apenko1985,Shapiro1986}. The basic
physical idea behind these works is that the effective potential seen by the
electrons in quantum Hall systems is quite smooth at the scale of the magnetic
length $l_{B}=\sqrt{\hbar c/|e|B}$ (here $\hbar$ is Planck's constant divided by
$2 \pi$, $c$ is the velocity of light, $B$ is the
magnetic field strength, and $|e|$ is the absolute value of the electric charge). This permits a simple mathematical treatment of the
Schr\"{o}dinger equation using as a small parameter the ratio of the
magnetic length to the typical correlation length of the random potential.
This point of view is certainly vindicated experimentally by the fact that the
impurities in semiconducting heterostructures are located outside the
two-dimensional layer of conduction electrons, while $l_B$ is an extremely
small length scale which falls below 10 nm for magnetic fields above 5 T.
It is tempting to believe that all aspects of the quantum Hall effect should be
captured accurately in this limit.

While the idea is certainly not new, it is interesting to note that no fully
quantum treatment of the high field regime is currently available, which
is the issue we want to address in this paper. Focusing first on the
semiclassical corrections to the $B=\infty$ limit, it is known that
systematic calculations are quite cumbersome, even at the lowest orders in
the $l_B$ expansion, due to the Landau level mixing \cite{Apenko1984,Haldane1997},
so that new tools are certainly needed. A first technical step in this direction was
made by two of us in a recent publication \cite{Champel2007} by introducing well-suited
coherent states Green's functions. These so-called vortex states with the quantum numbers
$\nu =(m,{\bf R})$, where $m$ is the Landau-level index, and ${\bf R}$ the position
of a localised vortexlike wave function, form an overcomplete basis of
eigenfunctions with no preferred symmetry, in contrast to the widely used
translation-invariant Landau states or the rotation-invariant circular states.
They thus permit a great adaptability to the spatial variations
of the local electric fields, coming from either random impurity donors,
confinement potentials due to external gates, or macroscopic voltage drops.
This formulation also allows one to classify Landau level mixing processes in a simple
and natural manner, an important point for the investigation of quantum transport
properties, as the matrix elements of the current density necessarily relate
adjacent Landau levels.

Our first implementation of this technique \cite{Champel2007} has demonstrated,
not surprisingly, that the usual semiclassical approach to the quantum Hall
effect (limited to spectral properties) could be easily recovered by a
straightforward expansion of the vortex Green's function in powers of the magnetic
length. In this view, the vortex state coordinate ${\bf R}$ can be identified in the limit $l_{B} \to 0$ with
the slow classical center-of-mass drifting motion, which completely decouples
from the faster cyclotron motion.

\subsection{Toward a unified quantum description at high magnetic field}
The present paper has two aims. First, we want to provide an accurate quantum
treatment of the local equilibrium properties of quantum Hall systems. For this
purpose, we offer simple functionals of the arbitrary local electrostatic
potential that describe both the local charge and current densities. These
results may have important bearings for microscopic modelizations of real
devices based on Hartree-Fock or more refined local density approximation (LDA) calculations, as they avoid
the numerical costs in solving the random Schr\"{o}dinger equation in a magnetic
field. The knowledge of the current density functional can be used in a second
step to obtain out of equilibrium transport equations, which take a simple
hydrodynamic form at high magnetic fields. This step allows us to derive
microscopically a simple and local expression for the conductivity tensor. We show that in
contrast to the well-known drift contribution to the transverse Hall
conductivity~\cite{Geller1994}, dissipative longitudinal components first appear at order
$l_B^2$. These contributions had not been obtained previously in the literature to our
knowledge. This method also allows one in principle to derive microscopically the dominant 
nonlocal corrections to local Ohm's law.
A general understanding of the dissipative features in the integer
regime seems now possible at the microscopic level.

\subsection{Organization of the paper}
Because the paper involves several novel technical developments, we hereafter guide the
reader through the main results obtained. Sec. \ref{section2} is used to
introduce  vortex Green's functions and reformulate in a more systematic manner
the results obtained in Ref. \onlinecite{Champel2007}. Two important formulas are found that
determine completely Green's functions both in the vortex coordinates [Eq.~(\ref{recursive})] and in the electronic coordinates [Eq.~(\ref{electronGreensimp})]. 
These are the starting point for the computation of all physical observables. 
The local electronic charge density is thus derived up to order $l_B^2$ in
Sec. \ref{section3}, and is given by formulas (\ref{electrondensity}),
(\ref{rho1eq}) and (\ref{rho2diag}). We emphasize beforehand that all these
expressions take into account quantum smearing effects from the wave function
and extend the semiclassical results (also derived in this section) to much
lower temperatures. 
Similarly, the (equilibrium) local electronic current density is computed to the same order in
Sec. \ref{section4}, and is given by Eqs.
(\ref{j0gen}), (\ref{j1fin}), (\ref{j1sub}), (\ref{j2nonlocal}), and (\ref{j3nonlocal}). Again, the semiclassical 
current density can be obtained from these expressions and is given by 
Eqs.~(\ref{j0}), (\ref{drift}), and (\ref{j0local})-(\ref{j3local}).
We then provide in Sec. \ref{section5} two important checks of our theory against an 
exactly solvable model of a one-dimensional parabolic confinement potential. First, the 
analytic semiclassical formulas for the local observables obtained in Secs. \ref{section3} 
and \ref{section4} are compared with the strict expansion in $l_B$ from the exact model 
and shown to match precisely, strengthening the mathematical foundation of our theory. 
Second, a more quantitative comparison is made between the quantum expressions obtained
for the electronic density in Sec. \ref{section3} and the exact results at finite 
values of the magnetic length. This shows that the expansion proposed here is converging 
quite rapidly, even for a confining potential that is not exceedingly smooth. 
Because the vortex states do not favor any special symmetric situation, similar quantitative 
results should be obtained for an arbitrary two-dimensional smooth disordered potential. 
In the limit of zero temperature, this comparison also shows the need for a
resummation of the quantum expressions, to infinite order in $l_B$, modifying
both the vortex wave functions and energies, and allowing a possible connection
to the edge state picture.
Finally, Sec. \ref{section6} investigates nonequilibrium properties in the integer
quantum Hall regime and provides a microscopic derivation of the local conductivity
tensor in the semiclassical regime [formula~(\ref{ohm})]. The origin of dissipation is discussed, and a general conclusion
showing future directions of our work closes the paper. Some extra technical
details are given in several appendixes.

\section{High field expansion within the vortex states representation
\label{section2}}

\subsection{Vortex states}

The vortex states under which our quantum high magnetic field theory reposes are
eigenstates of the free Hamiltonian
\begin{equation}
H_{0}=\frac{1}{2 m^{\ast}}\left( - i \hbar {\bm \nabla_{\bf r}} - \frac{e}{c}
{\bf A}({\bf r}) \right)^{2}
\label{H0},
\end{equation}
describing a single electron of effective mass $m^{\ast}$ and of charge $e=-|e|$
confined in a $(xy)$ two-dimensional plane under a perpendicular magnetic field
${\bf B}$ (pointing in the $z$ direction). In the symmetrical gauge
$$
{\bf A}({\bf r}) = \frac{\hbar c}{|e|} \frac{1}{2 l_{B}^{2}} \left(
\begin{array}{c}
-y \\ x
\end{array}
\right),
$$
the vortex wave functions, with quantum number $\nu=(m,{\bf R})$, are written in terms
of the electronic variables ${\bf r}$ as \cite{Apenko1983,Champel2007}
\begin{eqnarray}
\Psi_{m,{\bf R}}
({\bf r}) =
\frac{1}
{\sqrt{2 \pi m!}l_{B}}
\left(
\frac{x-X+i(y-Y)}{\sqrt{2}l_{B}} \right)^{m}
\nonumber \\
\times
\exp\left[
-\frac
{(x-X)^{2}+(y-Y)^{2}+2 i (yX-xY)}
{4 l_{B}^{2}}
\right].
\label{vortex}
\end{eqnarray}
The associated energy levels read
\begin{equation}
E_{m,{\bf R}} =\left(m+1/2 \right) \hbar \omega_{c} \equiv E_{m}
\label{energy},
\end{equation}
where $\omega_{c}=|e|B/m^{\ast}c$ is the cylclotron pulsation. The energy levels
are independent of the position ${\bf R}$, and their quantization is uniquely
related to the (topological) quantization of the circulation of any paths
enclosing the position ${\bf r}={\bf R}$ which corresponds to a phase
singularity of the wave function $\Psi_{m,{\bf R}}({\bf r})$ (note that this
wave function vanishes only at the point ${\bf r}={\bf R}$ where its phase is
ill-defined: It describes a vortex).

As is clear from Eq. (\ref{energy}) the Landau energy levels are highly degenerate,
so that there is a great freedom in the choice of a basis of states. However, a
judicious choice for the set of quantum numbers appears essential when
considering perturbations that lift this huge energy degeneracy.
A peculiarity of the vortex states, which could appear at the first glance as a
drawback, is that they are nonorthogonal with respect to the degeneracy quantum
number ${\bf R}$. Indeed, the overlap between two vortex states is given by
\begin{eqnarray}
\langle \nu_{1} | \nu_{2} \rangle & = & \delta_{m_{1},m_{2}} \,
\langle {\bf R}_{1} | {\bf R}_{2} \rangle\\
\langle {\bf R}_{1} | {\bf R}_{2} \rangle & = &
\exp\left[ -\frac{({\bf R}_{1}-{\bf R}_{2})^{2}-2i \hat{{\bf z}} \cdot
({\bf R}_{1}\times {\bf R}_{2})}{4 l_{B}^{2}} \right]
\end{eqnarray}
where $\hat{{\bf z}}$ is the unit vector along the perpendicular magnetic field.

On the contrary, the other well-known eigenstates of $H_{0}$, the Landau and
circular basis states, which are commonly used for quantum calculations, are
orthogonal. But since they are highly symmetric states, they lead to unsolved
technical difficulties when considering a random potential in high magnetic
fields, which mixes in a very complicated way the two quantum numbers. The vortex
basis, which has no intrinsic symmetry (the nonorthogonality of the vortex
states arises from this property), allows one to overcome this drawback. The
possibility \cite{Champel2007} to work with this basis is, in fact, provided by
the coherent character of the vortex position degree of freedom (the algebra
obeyed by the degeneracy quantum number ${\bf R}$ is that of coherent states).

\subsection{Dyson equation in the vortex representation}

From now on and throughout the paper, we consider that the Hamiltonian contains
in addition to the kinetic part $H_{0}$ a potential energy term $V({\bf r})$,
which we let completely unspecified
\begin{equation}
H=H_{0}+V(x,y).
\label{Hamiltonian}
\end{equation}
The Dyson equation written within the vortex representation $|\nu \rangle = |m, {\bf R}\rangle$ then takes the form
\cite{Champel2007}
\begin{equation}
(\omega-E_{m_{1}} \pm i \delta)G^{R,A}_{\nu_{1};\nu_{2}}(\omega)
=
\langle \nu_{1} | \nu_{2} \rangle
+\sum_{\nu_{3}} V_{\nu_{1};\nu_{3}} G^{R,A}_{\nu_{3};\nu_{2}}(\omega)
\label{Dyson},
\end{equation}
where $G^{R,A}_{\nu_{1};\nu_{2}}(\omega)$ are retarded and advanced Green's
functions connecting two vortex states $\nu_1$ and $\nu_2$ (in the energy
representation). The sum over the vortex quantum numbers $\nu $ appearing into
the Dyson equation stands for
\begin{equation}
\sum_{\nu}=\sum_{m=0}^{+ \infty} \int \!\!\! \frac{d{\bf R}}{2 \pi l_{B}^{2}}.
\end{equation}
The matrix elements of the potential $V({\bf r})$ in the vortex basis are given by
\begin{eqnarray}
V_{\nu_{1};\nu_{2}} &= &\int d^{2}{\bf r} \, V({\bf r}) \,
\Psi_{m_1,{\bf R}_{1}}^{\ast}({\bf r}) \,\Psi_{m_2,{\bf R}_{2}}({\bf r})
\nonumber \\
&= &\langle {\bf R}_{1} | {\bf R}_{2}\rangle \, v_{\nu_{1};\nu_{2}}
\label{vdef}
\end{eqnarray}
where, for a practical purpose which will appear obvious in the following,
the overlap between the two vortex states has been extracted.  Similarly
for retarded and advanced Green's function, we extract the vortices overlap
\begin{equation}
G^{R,A}_{\nu_{1};\nu_{2}}(\omega) =
\langle {\bf R}_{1} | {\bf R}_{2} \rangle
g^{R,A}_{\nu_{1};\nu_{2}},
\label{form}
\end{equation}
where the dependence on frequency $\omega$ is not
explicited anymore, in order not to burden the expressions.
Substituting expressions (\ref{vdef}) and (\ref{form}) in Eq. (\ref{Dyson}), we get a
Dyson equation for the function $g^{R,A}_{\nu_{1};\nu_{2}}$ which reads
\begin{eqnarray}
(\omega-E_{m_{1}} \pm i \delta)g^{R,A}_{\nu_{1};\nu_{2}}
& =&
\delta_{m_{1},m_{2}}
+\sum_{\nu_{3}} v_{\nu_{1};\nu_{3}} g^{R,A}_{\nu_{3};\nu_{2}}
\nonumber \\
&&
\times
\frac{
\langle {\bf R}_{1} | {\bf R}_{3} \rangle
\langle {\bf R}_{3} | {\bf R}_{2} \rangle
}{\langle {\bf R}_{1} | {\bf R}_{2} \rangle}
\label{Dyson2},
\end{eqnarray}
where
\begin{equation}
\frac{
\langle {\bf R}_{1} | {\bf R}_{3} \rangle
\langle {\bf R}_{3} | {\bf R}_{2} \rangle
}{\langle {\bf R}_{1} | {\bf R}_{2} \rangle}
=
\exp\left[
-\frac{
\left({\bf R}_{3}-({\bf c}_{12}+i{\bf d}_{12} \times \hat{{\bf z}})\right)^{2}
}{2 l_{B}^{2}}
\right]. \label{over}
\end{equation}
We have introduced here the center-of-mass coordinates ${\bf c}_{12}=({\bf R}_{1}+{\bf R}_{2})/2$
and the relative coordinates ${\bf d}_{12}=({\bf R}_{2}-{\bf R}_{1})/2$.

Provided that $V(x,y)$ is an analytic function of both $x$ and $y$,
the reduced matrix element $v_{\nu_{1};\nu_{2}}$ of the potential appearing
in Eq.~(\ref{vdef}) can be written \cite{Champel2007} as a series in powers of the magnetic length $l_{B}$
\begin{eqnarray}
v_{\nu_{1};\nu_{2}} & = & \sum_{j=0}^{+ \infty}
\left( \frac{l_{B}} {\sqrt{2}} \right)^{j}
v^{(j)}_{\nu_{1};\nu_{2}}
\label{v},\\
v^{(j)}_{\nu_{1};\nu_{2}} & = &
\sum_{k=0}^{j}
\frac{(m_{1}+k)!}{\sqrt{m_{1}!m_{2}!}}
\frac{\delta_{m_{1}+k,m_{2}+j-k}}{k!(j-k)!}
\nonumber \\
&& \times
(\partial_{X}+i \partial_{Y})^{k}
(\partial_{X}-i \partial_{Y})^{j-k}
\left.
V({\bf R})
\right|_{{\bf c}_{12}+i {\bf d}_{12}\times \hat{{\bf z}}}.
\nonumber \\
&&
\vspace*{-0.3cm}
\label{vj}
\end{eqnarray}

Solving exactly Dyson equation [Eq.~(\ref{Dyson2})] for an arbitrary potential $V$ is
certainly a formidable task. Remarkably, however, from the structure of this
equation, one can show that the function $g_{\nu_{1} ;\nu_{2}}$ depends on the two vortex
coordinates ${\bf R}_{1}$ and ${\bf R}_{2}$ through the special combination
${\bf c}_{12}+i{\bf d}_{12} \times \hat{{\bf z}}$ only.
Indeed, let us differentiate Eq.~(\ref{Dyson2}) with respect to the first vortex
position,
\begin{eqnarray}
(\omega-E_{m_{1}} \pm i \delta) (\partial_{X_{1}}- i \partial_{Y_{1}} ) g^{R,A}_{\nu_{1};\nu_{2}}
= \nonumber \\
\sum_{\nu_{3}}
(\partial_{X_{1}}- i \partial_{Y_{1}} )
\left\{
 v_{\nu_{1};\nu_{3}} g^{R,A}_{\nu_{3};\nu_{2}}
\frac{
\langle {\bf R}_{1} | {\bf R}_{3} \rangle
\langle {\bf R}_{3} | {\bf R}_{2} \rangle
}{\langle {\bf R}_{1} | {\bf R}_{2} \rangle} \right\}
\label{comb}.
\end{eqnarray}
Then, noting that from Eqs.~(\ref{over}) and (\ref{vj})
\begin{eqnarray}
(\partial_{X_{1}}- i \partial_{Y_{1}} )\left\{
\frac{
\langle {\bf R}_{1} | {\bf R}_{3} \rangle
\langle {\bf R}_{3} | {\bf R}_{2} \rangle
}{\langle {\bf R}_{1} | {\bf R}_{2} \rangle}
\right\}
&
=&
0 ,\\
(\partial_{X_{1}}- i \partial_{Y_{1}} )v_{\nu_{1};\nu_{3}}
&=&
0, \label{relav}
\end{eqnarray}
and considering Eq. (\ref{comb}), we arrive to the relation
\begin{equation}
\partial_{X_{1}}g^{R,A}_{\nu_{1};\nu_{2}}=i \partial_{Y_{1}}g^{R,A}_{\nu_{1};\nu_{2}} \label{rela1}.
\end{equation}
We can establish similarly from the other Dyson equation (i.e., $G=G_{0}+GVG_{0}$) that
\begin{equation}
\partial_{X_{2}}g^{R,A}_{\nu_{1};\nu_{2}}=-i \partial_{Y_{2}}g^{R,A}_{\nu_{1};\nu_{2}}. \label{rela2}
\end{equation}
We thus deduce from these two relations [Eqs. (\ref{rela1}) and (\ref{rela2})] that the
function $g$ depends on the vortex positions ${\bf R}_{1}$ and ${\bf R}_{2}$ in
the following way:
\begin{equation}
g_{\nu_{1} ;\nu_{2}}=g_{m_{1} ;m_{2}}\left(\frac{{\bf R}_{1}+{\bf R}_{2}+i({\bf R}_{2}-{\bf R}_{1}) \times \hat{{\bf z}}}{2} \right).
\label{dep}
\end{equation}
This exact result implies that vortex Green's functions will be entirely
determined once the function $g_{\nu_{1};\nu_{2}}$ at coinciding vortex
positions ${\bf R}_{1}={\bf R}_{2} \equiv {\bf R}$ are known (provided it is
analytic in the complex plane). This task is addressed in Sec. II C.

\subsection{High magnetic field expansion of vortex Green's function
\label{expand}
}

We are mainly interested in the high magnetic field regime, i.e., when the
magnetic length $l_{B}=\sqrt{\hbar c/|e|B}$ is small compared to the typical
length scale of the (possibly random) potential $V({\bf r})$. We aim at solving the Dyson
equation [Eq.~(\ref{Dyson2})] as a systematic expansion in powers of $l_{B}$, i.e., 
expanding the function $g_{\nu_{1};\nu_{2}}$ as
\begin{equation}
g_{\nu_{1};\nu_{2}} = \sum_{j=0}^{+\infty}
\left( \frac{l_{B}} {\sqrt{2}} \right)^{j}
g^{(j)}_{\nu_{1};\nu_{2}}.
\label{exp}
\end{equation}
This expansion is possible because, using the change in function (\ref{form}),
the nonanalytic dependence on the magnetic length $l_{B}$ which was contained
in the first term of the right-hand side (rhs) of Eq. (\ref{Dyson}) has been fully
transferred to the overlap terms [Eq. (\ref{over})] appearing in the integral
contribution of the Dyson equation [Eq.~(\ref{Dyson2})]. At large magnetic fields
(i.e., small $l_{B}$) and when ${\bf R}_{1}$ is close to ${\bf R}_{2}$ the main
contribution to the integral over ${\bf R}_{3}$ in Eq. (\ref{Dyson2}) comes when
${\bf R}_{3}$ is near both positions ${\bf R}_{1}$ and ${\bf R}_{2}$. Because Green's function~(\ref{dep}) depends on a linear combination of the two vortex
locations, it is enough to calculate vortex Green's function at coinciding
points ${\bf R}_{1} = {\bf R}_{2} \equiv {\bf R}$, so that, from
Eq.~(\ref{Dyson2}), only the value of the function $v_{m_{1},{\bf
R};m_{3},{\bf R}_{3}} g_{m_{3},{\bf R}_{3};m_{2},{\bf R}}$ has to be considered.
The Dyson equation can now be solved by expanding the nonlocal functions $g$
and $v$ around coinciding points using a Taylor series at ${\bf R}_{3}$ close to $
{\bf R}$,
\begin{widetext}
\begin{eqnarray}
g_{m_{3},{\bf R}_{3};m_{2},{\bf R}}
 & = & \sum_{k=0}^{+ \infty} \frac{\left[ (X_{3}-X)-i(Y_{3}-Y)\right]^{k}}{k! \, 2 ^{k}} \left(\partial_{X}+i\partial_{Y} \right)^{k} g_{m_{3};m_{2}}({\bf R}) \label{devg}, \\
v_{m_{1},{\bf R};m_{3},{\bf R}_{3}} &=&
\sum_{k'=0}^{+ \infty} \frac{\left[ (X_{3}-X)+i(Y_{3}-Y)\right]^{k'}}{k'! \, 2 ^{k'}} \left(\partial_{X}-i\partial_{Y} \right)^{k'} v_{m_{1};m_{3}}({\bf R})
\label{devv},
\end{eqnarray}
where we have taken into account the spatial dependences of $v$ [see Eq.
(\ref{vj})] and $g$ [see Eq. (\ref{dep})]. The integral over the vortex position
${\bf R}_{3}$ in Eq. (\ref{Dyson2}) can then be evaluated using the following property
of Gaussian integrals:
\begin{eqnarray}
\int \!\!\! \frac{d^{2} {\bf R}_{3}}{2 \pi l_{B}^{2}}
v_{m_{1},{\bf R};m_{3},{\bf R}_{3}} g_{m_{3},{\bf R}_{3};m_{2},{\bf R}}
\, e^{
-\frac{
\left({\bf R}_{3}-{\bf R}\right)^{2}
}{2 l_{B}^{2}} }
&= &
\int \!\!\! \frac{d^{2} {\bf R}_{3}}{2 \pi l_{B}^{2}} \sum_{k,k'=0}^{+\infty}
\left[(X_{3}-X)-i(Y_{3}-Y) \right]^{k} \left[(X_{3}-X)+i(Y_{3}-Y) \right]^{k'}
\nonumber \\
&&
\times \frac{e^{
-\frac{
\left({\bf R}_{3}-{\bf R}\right)^{2}
}{2 l_{B}^{2}} }}{k!k'! \, 2^{k+k'} }
 \left(\partial_{X}-i\partial_{Y} \right)^{k'} v_{m_{1};m_{3}}({\bf R})
 \left(\partial_{X}+i\partial_{Y} \right)^{k} g_{m_{3};m_{2}}({\bf R})
\\
& =&
\sum_{k=0}^{+ \infty}
\left(
\frac{
l_{B}}{\sqrt{2}}\right)^{2k} \frac{1}{k!}
 \left(\partial_{X}-i\partial_{Y} \right)^{k} v_{m_{1};m_{3}}({\bf R})
 \left(\partial_{X}+i\partial_{Y} \right)^{k} g_{m_{3};m_{2}}({\bf R})
.
\label{int}
\end{eqnarray}

Combining the different series expansions of the matrix elements of the
potential [Eqs. (\ref{v})-(\ref{vj})], of vortex Green's function $g^{R,A}$ [Eq.
(\ref{exp})], and of the integral term [Eq. (\ref{int})] in the Dyson equation,
the functions $g^{R,A}_{\nu_{1};\nu_{2}}$ at coinciding points ${\bf R}_{1}={\bf
R}_{2}$ are then entirely determined order by order in powers of the magnetic
length $l_{B}$. In fact, vortex Green's function $g^{(n)}$ at order $l_B^n$ is
related to the terms $g^{(l)}$ (with $l<n$) through
{\it a closed-form recursive relation}:
\begin{eqnarray}
g^{(n)}_{m_{1};m_{2}}({\bf R}) & = &
g^{(0)}_{m_{1}; m_{1}}({\bf R})
\sum_{l=0}^{n-1} \sum_{j=0}^{n-l} \sum_{k=0}^{(n-l)/2} \! \frac{1}{k!}
\delta_{n,2k+j+l} \!\! \sum_{m_{3}=m_{1}-j}^{m_{1}+j} \!\!
 \left(\partial_{X}+i\partial_{Y} \right)^{k} g^{(l)}_{m_{3};m_{2}}({\bf R})
 \left(\partial_{X}-i\partial_{Y} \right)^{k} v^{(j)}_{m_{1};m_{3}}({\bf R})
,
\label{recursive}
\end{eqnarray}
\end{widetext}
where the function $v^{(j)}$ is given by Eq.~(\ref{vj}) at coinciding
points
\begin{eqnarray}
v^{(j)}_{m_{1};m_{2}}({\bf R}) &= &
\sum_{k=0}^{j}
\frac{(m_{1}+k)!}{\sqrt{m_{1}!m_{2}!}}
\frac{\delta_{m_{1}+k,m_{2}+j-k}}{k!(j-k)!}
\nonumber \\
&& \times
(\partial_{X}+i \partial_{Y})^{k}
(\partial_{X}-i \partial_{Y})^{j-k}
V({\bf R}), \hspace*{0.6cm}
\label{vjbis}
\end{eqnarray}
and the zeroth order contribution $g^{(0)}$, that suffices to determine the
whole series, is given in Eq.~(\ref{g0}) below.

Obviously, the present method generates a {\em systematic} expansion for 
vortex Green's functions in series of the magnetic length. Even for a
disordered potential that is smooth on the scale of $l_B$, the question of the
accuracy and convergence of this expansion has to be addressed. We refer
the reader both to a general discussion of this important point in Sec.~\ref{remarks}
and to a concrete comparison with an exactly solvable model in Sec.~\ref{section5}.

\subsection{Vortex Green's functions up to order $l_B^3$}

For the calculations to follow in the rest of the paper, vortex Green's
functions up to order $l_B^3$ will be needed, and these useful expressions are
given here.
At leading order (zeroth order in magnetic length) the equation determining
the function $g^{(0)}$ is trivially found by setting $k=0$ in formula~(\ref{int}),
which is then reported in Dyson equation [Eq.~(\ref{Dyson2})],
\begin{equation}
(\omega-E_{m_{1}} \pm i \delta)
g^{(0)}_{m_{1};m_{2}}
({\bf R})
=
\delta_{m_{1},m_{2}}
+V\left({\bf R}
\right)
 g^{(0)}_{m_{1};m_{2}}
({\bf R})
\label{Dysong0}.
\end{equation}
This equation is entirely closed and yields straightforwardly
\begin{equation}
g^{(0) R,A}_{m_{1};m_{2}}({\bf R})
=
\frac{\delta_{m_{1},m_{2}}}{
\omega-\xi_{m_{1}}({\bf R}) \pm i \delta},
\label{g0}
\end{equation}
with $\xi_{m}({\bf R})=E_{m}+V({\bf R})$.
Green's function at leading order is diagonal with respect to the vortex
circulation quantum number $m$. We regard this robustness of $m$ independently
of the detailed form and strength of the potential $V$ as a signature of its
topological nature. We see that, in addition to a kinetic term ($E_{m}$), the
energy of the vortex state $\xi_{m}({\bf R})$  now also contains the value of the potential
energy $V({\bf R})$ at the vortex location, which lifts the huge degeneracy
of the Landau levels. This leading order of the calculation clearly corresponds to the
strict semiclassical limit~\cite{Trugman1983,Apenko1983,Apenko1984,Haldane1997} at $l_B=0$.

All subleading contributions are straightforwardly determined using the
recursive relation~(\ref{recursive}), which for $n=1$ gives the order
$l_B$ contribution
\begin{eqnarray}
g^{(1) R,A}_{m_{1};m_{2}}({\bf R})
 &= &
g^{(0)}_{m_{1};m_{1}}({\bf R}) \,
g^{(0)}_{m_{2};m_{2}}({\bf R}) \,
v^{(1)}_{m_{1};m_{2}}({\bf R})
\\
&=&
\frac{v^{(1)}_{m_{1};m_{2}}({\bf R})}{
\left(\omega-\xi_{m_{1}}({\bf R}) \pm i \delta \right)\left(\omega-\xi_{m_{2}}({\bf R}) \pm i \delta \right)
},
\nonumber \\
\vspace*{-0.3cm}
\label{g1}
\end{eqnarray}
where from equation~(\ref{vj})
\begin{eqnarray}
v^{(1)}_{m_{1};m_{2}}({\bf R}) &=&
\left[\sqrt{m_{2}} \delta_{m_{1}+1,m_{2}}\,(\partial_{X}+i\partial_{Y}) \right.
\nonumber \\
&&
\left.
+
\sqrt{m_{1}} \delta_{m_{1},m_{2}+1} \, (\partial_{X}-i\partial_{Y})
\right]V({\bf R}).
\nonumber \\
&& \vspace*{-0.3cm}
\end{eqnarray}
We thus see that a mixing between adjacent Landau levels appears in the
presence of a gradient of the potential $V$.

For the determination of the function $g^{(2)}$, the matrix elements of the
potential at order $l_B^2$ are needed, which read from~(\ref{vj})
\begin{widetext}
\begin{eqnarray}
v^{(2)}_{m_{1} ; m_{2}}({\bf R})
=
\frac{1}{2}
\frac{\mathrm{max}\left(m_{1},m_{2}\right)!}{\sqrt{m_{1}!m_{2}!}}
\left[
\delta_{m_{1}+2,m_{2}}
\left(\partial_{X}+i\partial_{Y} \right)^{2}
+
\delta_{m_{1},m_{2}+2}
\left(\partial_{X}-i\partial_{Y} \right)^{2} +
2(m_{1}+1) \delta_{m_{1},m_{2}} \Delta_{\bf R}
\right]V({\bf R})
.
\end{eqnarray}
Recursion relation~(\ref{recursive}) at this order gives
\begin{eqnarray}
g^{(2)}_{m_{1}; m_{2}}({\bf R}) = g^{(0)}_{m_{1}; m_{1}} \left[
v^{(2)}_{m_{1}; m_{2}}({\bf R})
g^{(0)}_{m_{2} ;m_{2}}({\bf R})
+\sum_{m_{3}}
v^{(1)}_{m_{1}; m_{3}}({\bf R})
g^{(1)}_{m_{3};m_{2}}({\bf R}) +
\delta_{m_{1},m_{2}}
{\bm \nabla}_{{\bf R}}V({\bf R})
\cdot
{\bm \nabla}_{{\bf R}}
 g^{(0)}_{m_{1};m_{1}}({\bf R})
\right]
.
\label{g2first}
\end{eqnarray}
The function $g^{(2)}$ consequently contains diagonal elements ($m_{1}=m_{2}$) and
elements mixing Landau levels separated by an energy of $2 \hbar \omega_{c}$ (terms
with $m_{1}=m_{2} \pm 2$),
\begin{eqnarray}
g^{(2)}_{m_{1};m_{2}}({\bf R})
&=& \delta_{m_{1},m_{2}}
\left[
(m_{1}+1)\frac{ \Delta_{\bf R} V}{\omega_{m_{1}}^{2}}
+
\left(\frac{m_{1}+1}{\omega_{m_{1}+1}} +
\frac{m_{1}}{\omega_{m_{1}-1}}
+
\frac{1}{\omega_{m_{1}}}
 \right)
\frac{\left| {\bm \nabla_{\bf R}} V \right|^{2}}{\omega_{m_{1}}^{2}}
\right]
\nonumber \\
&&
+
\sqrt{m_{1}+1} \sqrt{m_{1}+2} \,
\delta_{m_{1}+2,m_{2}}
\left[
\frac{(\partial_{X}+i\partial_{Y})^{2}V}{2 \omega_{m_{1}} \omega_{m_{2}}}
+
\frac{\left[(\partial_{X}+i\partial_{Y})V\right]^{2}}{\omega_{m_{1}}
\omega_{m_{1}+1}\omega_{m_{1}+2}}
\right]
\nonumber \\
&&
+
\sqrt{m_{2}+1} \sqrt{m_{2}+2} \,
\delta_{m_{1},m_{2}+2}
\left[
\frac{(\partial_{X}-i\partial_{Y})^{2}V}{2 \omega_{m_{1}} \omega_{m_{2}}}
+
\frac{\left[(\partial_{X}-i\partial_{Y})V\right]^{2}}{\omega_{m_{2}+2}
\omega_{m_{2}+1}\omega_{m_{2}}}
\right]
\label{g2},
\end{eqnarray}
\end{widetext}
 where we have introduced the short-hand notation
$\omega_{m}=\omega-\xi_{m}({\bf R})\pm i \delta$.

As for third-order Green's function, we shall not write here the full
expression, which is rather cumbersome. The derivation of these terms from Eq. (\ref{recursive}) is however
straightforward, and Appendix~\ref{apg3} provides
the components that are needed for subsequent calculations.

\subsection{Green's functions in the electronic representation}

The aim of this section is to connect local vortex Green's function, determined previously
in the  magnetic length expansion, to physical observables. For this purpose, we need
to express Green's functions in terms of the electronic positions ${\bf r}$,
which, thanks to the completeness relation satisfied by the vortex
states~\cite{Champel2007}, can be obtained as
\begin{equation}
G({\bf r},{\bf r}',\omega)= \sum_{\nu_{1}, \nu_{2}}
G_{\nu_{1};\nu_{2}}(\omega) \Psi_{\nu_{2}}^{\ast}({\bf r}') \Psi_{\nu_{1}}({\bf r}).
\end{equation}
Rewriting in terms of the vortex location ${\bf R}$ and circulation $m$, this expression reads
\begin{eqnarray}
G({\bf r},{\bf r}',\omega)
&=
&
 \int\!\!\! \frac{d^{2}{\bf R}_{1}}{2 \pi l_{B}^{2}}
\int\!\!\! \frac{d^{2}{\bf R}_{2}}{2 \pi l_{B}^{2}} \sum_{m_{1}, m_{2}}
 G_{m_{1}, {\bf R}_{1};m_{2},{\bf R}_{2}}(\omega)
\nonumber \\
&&
\times
 \Psi_{m_{2},{\bf R}_{2}}^{\ast}({\bf r}') \Psi_{m_{1},{\bf R}_{1}}({\bf r}). \label {electronGreen}
\end{eqnarray}

Besides the double integral in the above formula, the difficulty we immediately
encounter is that nonlocal vortex Green's function is in principle needed. Again, we
are going to see that the key formula~(\ref{dep}) allows us to reformulate this expression
in terms of local vortex Green's function determined in Sec.~\ref{expand}.
Inserting expression (\ref{form}), we first write the different exponential factors
appearing into the integrand of the
expression (\ref{electronGreen}) as
\begin{eqnarray}
\langle {\bf R}_{1} | {\bf R}_{2}\rangle \, e^{-\frac{({\bf r}-{\bf R}_{1})^{2}-2 i ({\bf r} \times {\bf R}_{1}) \cdot \hat{{\bf z}}}{4 l_{B}^{2}}}
 \, e^{-\frac{({\bf r}'-{\bf R}_{2})^{2}
+2 i ({\bf r}' \times {\bf R}_{2}) \cdot \hat{{\bf z}}
}{4 l_{B}^{2}}}
= \nonumber \\
e^{-\frac{2 {\bf d}_{12}^{2}}{l_{B}^{2}}}
 e^{-\frac{({\bf r}-{\bf R})^{2}-2 i ({\bf r} \times {\bf R}) \cdot \hat{{\bf z}}}{4 l_{B}^{2}}}
 \,
 e^{-\frac{({\bf r}'-{\bf R})^{2}
+2 i ({\bf r}' \times {\bf R}) \cdot \hat{{\bf z}}
}{4 l_{B}^{2}}},
\label{formexp}
\end{eqnarray}
where ${\bf R}={\bf c}_{12}-i {\bf d}_{12} \times \hat{{\bf z}}$ is a complex combination of the center-of-mass and of the relative vortex coordinates.
Similarly,
the polynomial parts of the vortex wave functions can be written as
\begin{eqnarray}
\left[x'-X_{2}-i(y'-Y_{2})\right]^{m_{2}}
\left[x-X_{1}+i(y-Y_{1})\right]^{m_{1}}
= \nonumber \\
\left[x'-X-i(y'-Y)\right]^{m_{2}}
\left[x-X+i(y-Y)\right]^{m_{1}}.
\label{pol}
\end{eqnarray}
It thus seems natural to introduce the change in variables $({\bf R}_{1},{\bf
R}_{2}) \to ({\bf R},{\bf d}_{12})$. The variables $X$ and $Y$ lie a priori on
lines in the complex plane as a result of the complex shift ($-i {\bf d}_{12}
\times \hat{{\bf z}}$). Using the analycity property of the functions in the
integrand, the contours of integration can be deformed to the real axes. The
dependences on the variables ${\bf R}$ and ${\bf d}_{12}$ in the function
$g_{m_{1},{\bf R}_{1};m_{2},{\bf R}_{2}}$ are made separable using Eq.
(\ref{dep}) and expanding the nonlocal $g$ function as

\begin{eqnarray}
g_{m_{1},{\bf R}_{1} ; m_{2},{\bf R}_{2}}
 &=& g_{m_{1}; m_{2}}({\bf c}_{12}+i {\bf d}_{12} \times \hat{{\bf z}})
\\
&=&
g_{m_{1}; m_{2}}({\bf R}+ 2 i {\bf d}_{12} \times \hat{{\bf z}})\\
 &=&
\sum_{j=0}^{+\infty} \frac{
\left[
2 i ({\bf d}_{12} \times \hat{{\bf z}}) \cdot {\bm \nabla}_{{\bf R}}
\right]^{j}}
{j!}
 g_{m_{1}; m_{2}}({\bf R})
\\
&=&
\sum_{j=0}^{+\infty} \sum_{k=0}^{j}
 \frac{
(d_{12 x}-i d_{12 y})^{k}(d_{12 x}+i d_{12 y})^{j-k}
}
{k!(j-k)! (-1)^{k} }
\nonumber \\
&& \times
(\partial_{X}+i \partial_{Y})^{k}(\partial_{X}-i \partial_{Y})^{j-k}
 g_{m_{1}; m_{2}}({\bf R}) \nonumber \\
&& \vspace*{-0.3cm}
\label{expg}
\end{eqnarray}
where we have used $2i({\bf d}_{12} \times \hat{{\bf z}}) \cdot {\bm \nabla_{\bf R}}=
(d_{12 x}+i d_{12 y})(\partial_{X}+i\partial_{Y})-(d_{12 x}-i d_{12
y})(\partial_{X}-i\partial_{Y})$ and then applied the binomial theorem.
Inserting expansion (\ref{expg}) into Eq. (\ref{electronGreen}) and using Eqs. (\ref{formexp})-(\ref{pol}), we can then perform the integral over the
relative coordinates ${\bf d}_{12}$ to finally obtain
\begin{eqnarray}
G({\bf r},{\bf r}',\omega)= \int\!\!\! \frac{d^{2}{\bf R}}{2 \pi l_{B}^{2}}
 \sum_{m, m'}
\Psi_{m',{\bf R}}^{\ast}({\bf r}') \Psi_{m,{\bf R}}({\bf r})
\nonumber
\\
\times
\sum_{k=0}^{+ \infty} \frac{1}{k!} \,
\left(-\frac{l_{B}^2}{2} \Delta_{\bf R} \right)^{k}
g_{m;m'}({\bf R})
. \label {electronGreensimp}
\end{eqnarray}

\subsection{On the convergence of the $l_B$ expansion}
\label{remarks}

Equation~(\ref{electronGreensimp}) above is clearly remarkable as it connects local vortex Green's
function $g_{m;m'}({\bf R})$ to the nonlocal electronic propagator
$G({\bf r},{\bf r}',\omega)$, from which all equilibrium physical properties can
be obtained. Because the vortex wave functions appearing in this expression have a finite
extension in space of order $l_B$, the combination of Eq.~(\ref{electronGreensimp})
with recursion relation~(\ref{recursive}), which encodes the small $l_B$ expansion
of $g_{m;m'}({\bf R})$, allows one to systematically obtain {\it quantum} expressions for
the physical observables, i.e., that are naively valid at small but finite magnetic length.
In contrast, the usual semiclassical expansion
\cite{Apenko1984,Geller1994,Haldane1997} is formulated in a strict
$l_B\rightarrow0$ limit, which would appear in our formalism as a further expansion
in powers of $l_B$ of the wave functions in Eq.~(\ref{electronGreensimp}). The latter semiclassical
expansion, which is analyzed in detail in Secs~\ref{semirho} and \ref{semij},
is clearly asymptotic in nature and certainly fails to be accurate at
low temperature, where quantum effects set in (this is explicitely demonstrated
in Sec.~\ref{section5} with the comparison to an exactly solvable model).

A central question is whether our expansion, performed order by order in powers of
$l_B$ for vortex Green's function $g({\bf R})$, does fully capture the quantum
effects that survive at small but nonzero magnetic length. As discussed by
several authors~\cite{Trugman1983,Joynt}, the Schr\"{o}dinger equation becomes
integrable in this limit, with constants of motion associated to equipotential
lines of the random potential. The wave functions are then localized on a scale
$l_B$ transverse to these equipotentials and spread on the whole constant
energy contour. While this picture is certainly appealing, difficulties arise
for generating a systematic expansion that takes into account Landau level mixing.
In contrast, the vortex states are flexible enough to capture these important
contributions but correspond to a starting point where wave functions are fully
localized at the scale $l_B$.
Indeed, the first term in our expansion [Eq.~(\ref{g0})] indicates
that eigenstates in the $B=\infty$ limit correspond to equipotential points, not
lines, and this is related to the fact that the vortex wave functions, which form an
overcomplete basis, become pointlike and thus orthogonal in this limit.
Physically, one expects that quantum fluctuations will play a crucial role as soon as
$B$ is finite by selecting orthogonal, and therefore more extended, wave packets.
Mathematically, this phenomenon is reflected in our formalism by the presence of terms
that have to be kept at each order of the small $l_B$ expansion for vortex Green's
functions [e.g., at order $l_{B}^{2}$, these are the terms with a single frequency $\omega_m$ in
Eq.~(\ref{g2})].
The need for resumming this expansion comes in fact from the Taylor expansion
to finite order of  nonlocal vortex Green's function [Eqs.~(\ref{devg}) and (\ref{expg})].
Fortunately, as terms at an arbitrary order in $l_B$ can be generated through 
relation~(\ref{recursive}), it is possible to achieve a resummation of the leading
contributions to vortex Green's function. Although these considerations are beyond the
scope of the present paper, a first step in this direction is presented for the case of the local
electronic density in Sec.~\ref{resum}.

\section{Electron density \label{section3}}

\subsection{General expression}
Vortex Green's functions being determined, one can then derive quantum microscopic expressions for the local physical observables.
The equilibrium local density is related to distribution (lesser
component) Green's function $G^{<}$ in the electronic representation by the
general formula
\begin{eqnarray}
\rho({\bf r}) & = & - i \int \!\!\! \frac{d \omega}{2 \pi} G^{<}({\bf r},{\bf r},\omega)
\label{densitydef}\\
& = & - i \int \!\!\! \frac{d \omega}{2 \pi}
\int\!\!\! \frac{d^{2}{\bf R}}{2 \pi l_{B}^{2}}
 \sum_{m, m'}
\Psi_{m',{\bf R}}^{\ast}({\bf r}) \Psi_{m,{\bf R}}({\bf r})
\nonumber \\
&& \times
\sum_{k=0}^{+ \infty} \frac{1}{k!} \,
\left(-\frac{l_{B}^2}{2} \Delta_{\bf R} \right)^{k}
g^<_{m;m'}({\bf R}),
\label{densitystart}
\end{eqnarray}
where Eq.~(\ref{electronGreensimp}) has been used. The distribution
function in the vortex basis reads  at equilibrium
\begin{eqnarray}
-i g^<_{m;m'}({\bf R}) & = & i n_{F}(\omega)\left(
g^R_{m;m'}({\bf R}) - g^A_{m;m'}({\bf R}) \right)
 \label{electronGreenlesser},
\\
n_{F}(\omega) & = & \frac{1}{1+\exp\left[(\omega-\mu^{\ast})/T \right]},
\end{eqnarray}
where the Fermi-Dirac distribution function has been introduced, with $\mu^{\ast}/e=\Phi_{0}$
the electrochemical potential (which is constant in space at thermodynamic equilibrium).

The computation of the electronic density at an arbitrary order in the magnetic
length expansion is now straightforward using Eq.~(\ref{recursive}) to
generate successive contributions to $g^<_{m;m'}({\bf R})$. An important
remark is however in order. While equation~(\ref{densitystart}) involves
local Green's function relative to the vortex position ${\bf R}$, it takes into
account all Landau level mixing processes (terms with $m\neq m'$). As we
will discover in the following calculations, the combination of vortex wave functions
$ \Psi_{m',{\bf R}}^{\ast}({\bf r}) \Psi_{m,{\bf R}}({\bf r})$ involves an extra
power $l_B^{|m-m'|}$. For this reason, the contribution from $g^{(1)}$, which couples
adjacent Landau levels, is actually of order $l_B^2$ and not $l_B$. Similarly,
the diagonal ($m=m'$) terms in $g^{(2)}$ are indeed of order $l_B^2$, while
contributions with $m=m'\pm2$ [see Eq.~(\ref{g2})] are overally of order $l_B^4$ and will be discarded in the following.

\subsection{Electron density at leading order}
At leading order of the expansion in $l_{B}$, vortex Green's function
is given by Eq.~(\ref{g0}) so that the distribution function reads
\begin{equation}
-i g^{(0) <}_{m;m'}({\bf R}) =
2\pi n_{F}(\omega)\, \delta_{m,m'}\, \delta( \omega-\xi_{m}({\bf R}))
.
\end{equation}
Inserting this in Eq.~(\ref{densitystart}) and performing the frequency sum, we
obtain the local electron density
\begin{eqnarray}
\rho^{(0)}({\bf r}) & = & \int\!\!\! \frac{d^{2}{\bf R}}{2 \pi l_{B}^{2}}
 \sum_{m=0}^{+ \infty} \left| \Psi_{m,{\bf R}}({\bf r}) \right|^{2}
n_{F}\left( \xi_{m}({\bf R})\right) 
\label{electrondensity}\\
|\Psi_{m,{\bf R}}({\bf r})|^2 & = &
\frac{1}{2 \pi m!l_{B}^2}
\left|\frac{{\bf R}-{\bf r}}{\sqrt{2} l_{B}}\right|^{2m}
\!\!\!\!
\exp\left[ -\frac{({\bf R}-{\bf r})^2}{2 l_{B}^{2}}
\right]
.
\nonumber \\
&& \vspace*{-0.3cm}
\label{norm}
\end{eqnarray}
We note that this zeroth order contribution (\ref{electrondensity}) is already more powerful than the
expression for the electron density that is obtained in the strict limit $l_{B} \to
0$ of infinite magnetic field,
\begin{equation}
\rho^{(0)}({\bf r})\rightarrow \frac{1}{2 \pi l_{B}^{2}}
 \sum_{m=0}^{+ \infty}
n_{F}\left( \xi_{m}({\bf r})\right)
 \label{electrondensity2}.
\end{equation}
This semiclassical result (\ref{electrondensity2}) has been widely used in the
literature~\cite{Chklovskii1992,Geller1994} as a basis to screening
calculations. It however ignores the fact that the physical density cannot vary faster
than the scale $l_B$ as is clear from Eqs.~(\ref{electrondensity}) and~(\ref{norm}),
and this leads to important quantitative differences, especially at low temperatures
where quantum smearing effects supersede the thermal broadening of the
density~\cite{Siddiki2004}.

Thus, expression (\ref{electrondensity}) clearly includes important resummations
of a purely semiclassical expansion of the physical density such
as Eq.~(\ref{electrondensity2}), which is naturally encoded order by order
in the expansion of vortex Green's functions. Before addressing the question of
the convergence of both types of calculations in Secs~\ref{semirho} and \ref{semij},
we compute now the next order contribution to the density.

\subsection{Electron density at order $l_B^2$}

As mentioned above, the contribution of order $l_B^2$ to the density comes
from three origins: The non-diagonal part of $g^{(1)}$, the diagonal part of
$g^{(2)}$, and the term $\Delta_{\bf R} g^{(0)}$ in expression~(\ref{densitystart}) appearing with $k=1$.
Let us investigate these different contributions in turn.

\subsubsection{Contribution from $g^{(1)}$}
The contribution from $g^{(1)}$ is obtained by inserting Eq.~(\ref{g1}) in Eq.~(\ref{exp}),
and reporting vortex Green's function in Eq.~(\ref{densitystart}):
\begin{widetext}
\begin{eqnarray}
\rho^{(1)}({\bf r}) & = &
\int\!\!\! \frac{d^{2}{\bf R}}{2 \pi l_{B}^{2}} \sum_{m=0}^{+ \infty}
 \frac{l_{B}}{\sqrt{2}} \sqrt{m+1}
\left[
\Psi_{m+1,{\bf R}}^{\ast}({\bf r}) \Psi_{m,{\bf R}}({\bf r})(\partial_{X}V+i\partial_{Y}V)
+
\Psi_{m,{\bf R}}^{\ast}({\bf r}) \Psi_{m+1,{\bf R}}({\bf r})(\partial_{X}V-i\partial_{Y}V)
\right]
\nonumber
\\
&&
\times
\frac{
 \left[ n_{F}\left(\xi_{m+1}({\bf R})\right)
-n_{F}\left(\xi_{m}({\bf R})\right)
\right]
}{\hbar \omega_{c}}.
\label{ligne1}
\end{eqnarray}
\end{widetext}
It is useful to note the relation, proved in the Appendix \ref{approofs},
\begin{eqnarray}
\sqrt{m+1}
\,
\Psi_{m+1, {\bf R}}^{\ast}({\bf r}) \Psi_{m,{\bf R}} ({\bf r})
= \nonumber \\
-
\frac{l_{B}}{\sqrt{2}}
\left( \partial_{x}-i\partial_{y}
\right)
\sum_{p=0}^{m} \left|\Psi_{p,{\bf R}}({\bf r}) \right|^{2}
\label{practical}
\end{eqnarray}
which shows that the product of wave functions $\Psi_{m+1}^{\ast}\Psi_{m}$
with adjacent Landau indices generates terms that behave as $l_{B}$, so that
the contribution from $g^{(1)}$ to the density is indeed of order $l_B^2$.
Using Eq. (\ref{practical}), the expression (\ref{ligne1}) can be written in the
equivalent form
\begin{eqnarray}
\rho^{(1)}({\bf r}) &=&
-
\int\!\!\! \frac{d^{2}{\bf R}}{2 \pi l_{B}^{2}} \sum_{m=0}^{+ \infty}
\frac{
 \left[ n_{F}\left(\xi_{m+1}({\bf R})\right)
-n_{F}\left(\xi_{m}({\bf R})\right)
\right]
}{\hbar \omega_{c}}
\nonumber \\
&&
\times
l_{B}^{2}
{\bm \nabla}_{{\bf R}} V
\cdot
\sum_{p=0}^{m}
{\bm \nabla}_{{\bf r}}
\left|
\Psi_{p,{\bf R}}({\bf r})
\right|^{2} \label{utile} .
\end{eqnarray}
Performing one of the discrete sums and  an integration by parts and
noting that ${\bm \nabla}_{{\bf r}}
\left|
\Psi_{m,{\bf R}}({\bf r})
\right|^{2}=- {\bm \nabla}_{{\bf R}}
\left|
\Psi_{m,{\bf R}}({\bf r})
\right|^{2}$,
we end up with
\begin{widetext}
\begin{equation}
\rho^{(1)}({\bf r}) =
\int\!\!\! \frac{d^{2}{\bf R}}{2 \pi l_{B}^{2}}
 \sum_{m=0}^{+ \infty}
\left| \Psi_{m,{\bf R}}({\bf r}) \right|^{2}
l_{B}^{2}
\left(
n'_{F}\left( \xi_{m}({\bf R})\right)
\frac{\left|{\bm \nabla_{\bf R}} V \right|^{2}}{\hbar \omega_{c}}
+ n_{F}\left( \xi_{m}({\bf R})\right)
\frac{\Delta_{\bf R} V}{\hbar \omega_c}
\right)
.
\label{rho1eq}
\end{equation}

\subsubsection{Contribution from $g^{(2)}$}
From Eq.~(\ref{densitystart}) the remaining contributions to the electron density at order $l_B^2$
are clearly
\begin{equation}
\rho^{(2)}({\bf r}) =
 - i \int \!\!\! \frac{d \omega}{2 \pi}
\int\!\!\! \frac{d^{2}{\bf R}}{2 \pi l_{B}^{2}}
 \sum_{m}
\Psi_{m,{\bf R}}^{\ast}({\bf r}) \Psi_{m,{\bf R}}({\bf r})
\left[
\frac{l_{B}^2}{2} g^{(2)<}_{m;m}({\bf R})
-\frac{l_{B}^2}{2} \Delta_{\bf R}
g^{(0)<}_{m;m}({\bf R})\right].
\end{equation}
Using results (\ref{g0}) and (\ref{g2}), we obtain
\begin{eqnarray}
\rho^{(2)}({\bf r} )
&=&
\int\!\!\! \frac{d^{2}{\bf R}}{2 \pi l_{B}^{2}}
 \sum_{m=0}^{+ \infty}
\left|
\Psi_{m,{\bf R}}({\bf r})
\right|^{2}
\frac{l_{B}^{2}}{2}
\left(
n'_{F}\left( \xi_{m}({\bf R})\right)
\left[
m
 \Delta_{\bf R} V
- \frac{\left|{\bm \nabla_{\bf R}} V \right|^{2}}{\hbar \omega_{c}}
\right]
-n_{F}''\left( \xi_{m}({\bf R})\right) \frac{\left|{\bm \nabla_{\bf R}} V \right|^{2}}{2}
\nonumber
\right.\\
&&
\left.
+ \frac{\left|{\bm \nabla_{\bf R}} V \right|^{2}}{(\hbar \omega_{c})^{2}} \left[
(m+1) n_{F}\left( \xi_{m+1}({\bf R})\right)
-(2m+1) n_{F}\left( \xi_{m}({\bf R})\right)
+ m \, n_{F}\left( \xi_{m-1}({\bf R})\right)
\right]
\right).
\label{rho2diag}
\end{eqnarray}
\end{widetext}
The components $\delta_{m,m'\pm 2}$ of the function $g^{(2)}$ [see Eq. (\ref{g2})] have not
been included in the calculations since they generate corrections to the density of the order $l_B^4$.

The final results for the electronic density up to order $l_B^2$, given in
formulas~(\ref{electrondensity}), (\ref{rho1eq}) and (\ref{rho2diag}), will be exploited in
detail by a comparison with an exactly solvable model in Sec.~\ref{section5}.
In anticipation to Sec.~\ref{resum}, we note that all these expressions
require at very low temperature a resummation procedure, which leads to define
renormalized energies and wave functions.

\subsection{Semiclassical density: The strict $l_B\rightarrow0$ expansion}
\label{semirho}

As already mentioned above, it is possible to express the electron density
under the form of a strict expansion in powers of the magnetic length. This
corresponds exactly to a systematic semiclassical expansion with respect to
the center-of-mass motion (the orbital motion giving rise to the Landau levels
is always treated quantum mechanically).

In this section, we write down explicitly the first corrections to the well-known
semiclassical expression, i.e., Eq. (\ref{electrondensity2}), for the electron
density. Nonlocal expressions (\ref{electrondensity}), (\ref{rho1eq}) and (\ref{rho2diag}) can be transformed into local ones, in a similar way as has been done at the level of the Dyson equation [Eq. (\ref{Dyson2})], by replacing any integral over the vortex position ${\bf R}$ in the following way:

\begin{widetext}
\begin{eqnarray}
\int\!\!\!
 d^{2}{\bf R}
 \left| \Psi_{m,{\bf R}}({\bf r}) \right|^{2}
f({\bf R})
&=&
\int\!\!\! d^{2}{\bf R}
 \left| \Psi_{m,{\bf R}}({\bf 0}) \right|^{2} f({\bf r}+{\bf R})
=
\int\!\!\! d^{2}{\bf R}
 \left| \Psi_{m,{\bf R}}({\bf 0}) \right|^{2}
\sum_{j=0}^{+\infty}
\frac{
\left({\bf R} \cdot {\bm \nabla}_{{\bf r}} \right)^{j} }{j!}f({\bf r})
\\
&=&
\sum_{j=0}^{+\infty}
\frac{(m+j)!}{(j!)^{2} m!} \left(\frac{l_{B}^2}{2} \Delta_{{\bf r}} \right)^{j} f({\bf r}),
\label{general}
\end{eqnarray}
where $f({\bf R})$ represents an arbitrary function of the vortex position.
Using Eq. (\ref{general}) in Eq.~(\ref{electrondensity}), we get the semiclassical contribution of order $l_B^2$ arising from the
expansion of $\rho^{(0)}$,
\begin{eqnarray}
\rho^{(0)}({\bf r}) =
\frac{1}{2 \pi l_{B}^{2}}
\sum_{m=0}^{+ \infty}
\left[
n_{F}\left( \xi_{m}({\bf r})\right)+
\frac{l_{B}^{2}}{2}(m+1) \Delta_{{\bf r}}
n_{F}\left( \xi_{m}({\bf r})\right)
\right].
\end{eqnarray}
At order $l_B^2$, the contribution from $\rho^{(1)}$ [Eq. (\ref{rho1eq})]
is readily obtained as
\begin{equation}
\rho^{(1)}({\bf r}) =
\frac{1}{2 \pi l_B^2} \sum_{m=0}^{+ \infty} l_B^2
\left(
n'_{F}\left( \xi_{m}({\bf r})\right)
\frac{\left|{\bm \nabla_{\bf r}} V \right|^{2}}{\hbar \omega_{c}}
+ n_{F}\left( \xi_{m}({\bf r})\right)
\Delta_{\bf r} V
\right)
.
\label{rho1bis}
\end{equation}
Similarly, the contribution from $\rho^{(2)}$ [Eq. (\ref{rho2diag})] reads
after simplification
\begin{eqnarray}
\rho^{(2)}({\bf r} )=
\frac{1}{2 \pi l_{B}^{2}}
 \sum_{m=0}^{+ \infty}
\frac{l_{B}^{2}}{2}
\left(
n'_{F}\left( \xi_{m}({\bf r})\right)
\left[
m \Delta_{\bf r} V
- \frac{\left|{\bm \nabla_{\bf r}} V \right|^{2}}{\hbar \omega_{c}}
\right]
-n_{F}''\left( \xi_{m}({\bf r})\right)
\frac{\left|{\bm \nabla_{\bf r}} V \right|^{2}}{2}
\right).
\end{eqnarray}
Collecting all the different contributions, we find that the total electronic density
in high magnetic fields including the first quantum corrections of order $l_{B}^{2}$ is given by
\begin{eqnarray}
\rho({\bf r}) & = &
\frac{1}{2 \pi l_{B}^{2}}
 \sum_{m=0}^{+ \infty}
n_{F}\left( \xi_{m}({\bf r})\right)
+
\frac{1}{2 \pi l_{B}^{2}}
 \sum_{m=0}^{+ \infty}
l_{B}^{2}
\Bigg[
n_{F}\left( \xi_{m}({\bf r})\right)
\frac{\Delta_{\bf r} V}{\hbar \omega_{c}}
+\left(m+\frac{1}{2} \right) n'_{F}\left( \xi_{m}({\bf r})\right) \Delta_{\bf r} V
\nonumber \\
& & +\frac{1}{2} n'_{F}\left( \xi_{m}({\bf r})\right)\frac{\left|{\bm \nabla_{\bf r}} V \right|^{2}}{\hbar \omega_{c}}
+\frac{1}{2}
\left(m+\frac{1}{2} \right) n''_{F}\left( \xi_{m}({\bf r})\right)
\left|{\bm \nabla_{\bf r}} V \right|^{2}
\Bigg].
\label{densitysecondorder}
\end{eqnarray}

In order to physically interpret this result, we can alternatively write
\begin{equation}
\rho({\bf r}) =
\frac{1}{2 \pi l_{B}^{2}}
 \sum_{m=0}^{+ \infty} \left[
n_{F}(\widetilde{\xi}_m({\bf r}))
+ l_{B}^{2}
n_{F}\left( \xi_{m}({\bf r})\right)
\frac{\Delta_{\bf r} V}{\hbar \omega_{c}}
+l_B^2 n_{F}'\left( \xi_{m}({\bf r})\right)
\frac{\left|{\bm \nabla_{\bf r}} V \right|^{2}}{\hbar \omega_{c}}
+\frac{l_B^2}{2}
\left(m+\frac{1}{2} \right) \Delta_{\bf r} n_{F}\left( \xi_{m}({\bf r})\right)
\right],
\label{rhofinal}
\end{equation}
\end{widetext}
where corrections proportional to $n_F'$ in $\rho^{(2)}$ have been absorbed into
renormalized Landau-level energies as
\begin{equation}
\widetilde{\xi}_m({\bf r}) = \xi_{m}({\bf r})
+\frac{l_B^2}{2}\left(m+\frac{1}{2} \right) \Delta_{\bf r} V
-\frac{l_B^2}{2} \frac{\left|{\bm \nabla_{\bf r}} V \right|^{2}}{\hbar
\omega_{c}}.
\label{Eshift}
\end{equation}
These semiclassical energies [Eq.~(\ref{Eshift})] have been previously found in the
literature~\cite{Apenko1984,Haldane1997} with techniques based on effective
Hamiltonians, which neglect Landau level mixing, and thus do not allow one to compute
the full local-density expression ~(\ref{rhofinal}).
The second term appearing in the rhs of Eq.~(\ref{rhofinal}), proportional to
$ l_{B}^{2} n_{F} \Delta_{\bf r} V/\hbar \omega_{c}$, reflects the small
but nonzero compressibility of the electron gas even in the absence of
electron-electron interaction. This term and also the third term
in Eq.~(\ref{rhofinal}), both derived from $g^{(1)}$, stem from adjacent Landau level mixing
processes and can be interpreted as small corrections to the wave function for
a smooth potential $V({\bf r})$.
Clearly, the fourth term in Eq.~(\ref{rhofinal}), proportional to $l_B^2
\Delta_{\bf r} n_F$, is only a small correction if the electronic density is
also smooth at the scale of $l_B$. This indicates that the semiclassical picture
breaks down at low temperature. In this case, one has to resort to fully quantum
expressions such as Eqs.~(\ref{electrondensity}), (\ref{rho1eq}) and (\ref{rho2diag}), as will be 
discussed in Sec.~\ref{comp}.

\subsection{Electron-electron interactions and screening}

As a result of electron-electron interactions, the potential $V$ entering into the
previous expressions through the Fermi function is possibly very different from the 
bare electrostatic potential (related to confining gates and random impurities outside the
two-dimensional electron gas), and has to be determined self-consistently from screening
theory. Previous work, rooted in the semiclassical picture, used expressions
for the electron density such as Eq.~(\ref{electrondensity2}) as a starting point
for Thomas-Fermi type of calculations 
\cite{Chklovskii1992,Chklovskii1993,Lier1994}. The physical picture that
emerged from these studies is that the sample separates into either compressible 
regions, where screening of the bare potential is almost perfect and the
electronic density varies spatially, or into incompressible regions, where the 
density is almost exactly pinned and the gradient of the effective potential 
is nonzero. Further work~\cite{Siddiki2004} has however shown that important
deviations result from a better resolution of the self-consistent problem
within a Hartree approximation that includes quantum smearing effects from
the electronic wave functions. Most of these calculations are performed in
simplified one-dimensional geometries, since the self-consistent resolution of the
Schr\"{o}dinger equation becomes prohibitive for an arbitrary disorder
landscape~\cite{Wulf1988,Sohrmann2007}. One can hope that our
high-field expression for the density will turn out to be a very useful tool in
this context of the study of electron-electron interaction effects in a disordered system.

\section{Electron current density \label{section4}}

\subsection{General expression}

The local electron current density is defined in terms of electronic Green's
function by
\begin{equation}
{\bf j}({\bf r},\omega)
=
\left.
\left[
\frac{e \hbar}{2 m^{\ast}} ({\bm \nabla}_{{\bf r}'}-{\bm \nabla}_{{\bf r}})
+ i \frac{e^{2}}{m^{\ast}c} {\bf A}
\right]
 G^{<}({\bf r},{\bf r}',\omega) \right|_{{\bf r}'={\bf r}}. \label{electronspectral}
\end{equation}
In a first step, this expression can be written in terms of vortex Green's
functions. Inserting expression~(\ref{electronGreensimp}), we get
\begin{widetext}
\begin{eqnarray}
{\bf j}({\bf r},\omega)
=
\frac{e \hbar}{2 m^{\ast}}
 \int\!\!\! \frac{d^{2}{\bf R}}{2 \pi l_{B}^{2}}
\sum_{m, m'}
\left[
 \Psi_{m,{\bf R}}({\bf r}) {\bm \nabla}_{{\bf r}} \Psi_{m',{\bf R}}^{\ast}({\bf r})
-
 \Psi_{m',{\bf R}}^{\ast}({\bf r}) {\bm \nabla}_{{\bf r}} \Psi_{m,{\bf R}}({\bf r})
+ 2 i \frac{e}{\hbar c} {\bf A} \Psi_{m',{\bf R}}^{\ast}({\bf r}) \Psi_{m,{\bf R}}({\bf r})
\right]
\nonumber \\
\times
\sum_{k=0}^{+ \infty} \frac{1}{k!} \,
\left(-\frac{l_{B}^2}{2} \Delta_{\bf R} \right)^{k}
g_{m, ;m'}^{<}({\bf R},\omega)
.
\label{j}
\end{eqnarray}
The dependence on the variable ${\bf r}$ is contained only in the wave functions
and the vector potential ${\bf A}$, which are all known (we remind that we have
chosen the symmetrical gauge to write down explicitly the vortex wave functions).
Using the relation
\begin{equation}
{\bm \nabla}_{{\bf r}} \Psi_{m}=
\left(
\begin{array}{c}
\frac{\sqrt{m}}{\sqrt{2}l_{B}} \Psi_{m-1}
-\frac{\sqrt{m+1}}{\sqrt{2}l_{B}} \Psi_{m+1}
+\frac{iy}{2 l_{B}^{2}} \Psi_{m}
\\
i \frac{\sqrt{m}}{\sqrt{2}l_{B}} \Psi_{m-1}
+i\frac{\sqrt{m+1}}{\sqrt{2}l_{B}} \Psi_{m+1}
-\frac{ix}{2 l_{B}^{2}} \Psi_{m}
\end{array}
\right),
\label{60}
\end{equation}
we can rewrite the bracketed term in~(\ref{j}) as
\begin{eqnarray}
i {\bm \nabla}_{{\bf r}}
\left[
\Psi_{m,{\bf R}}({\bf r}) \Psi_{m',{\bf R}}^{\ast}({\bf r})
\right] \times \hat{{\bf z}}
-\frac{\sqrt{2}}{l_{B}}
\left(
\begin{array}{c}
\sqrt{m'+1} \Psi_{m, {\bf R}}({\bf r}) \Psi^{\ast}_{m'+1, {\bf R}}({\bf r})-\sqrt{m+1} \Psi_{m+1, {\bf R}}({\bf r}) \Psi^{\ast}_{m', {\bf R}}({\bf r})
\\
i\sqrt{m'+1} \Psi_{m, {\bf R}}({\bf r}) \Psi^{\ast}_{m'+1, {\bf R}}({\bf r})+i\sqrt{m+1} \Psi_{m+1, {\bf R}}({\bf r}) \Psi^{\ast}_{m', {\bf R}}({\bf r})
\end{array}
\right)
.
\label{yes}
\end{eqnarray}
Inserting expression (\ref{yes}) in formula (\ref{j}), we finally get our
starting point for the computation of the current density,
\begin{eqnarray}
{\bf j}({\bf r},\omega)
=\frac{e \hbar}{2m^{\ast}}
\Bigg[
\hat{{\bf z}} \times {\bm \nabla_{\bf r}} \rho({\bf r},\omega)
-\frac{\sqrt{2}}{l_{B}}
 \int\!\!\! \frac{d^{2}{\bf R}}{2 \pi l_{B}^{2}} \sum_{m, m'}
\left(
\begin{array}{c}
\sqrt{m'+1} \Psi_{m, {\bf R}}({\bf r}) \Psi^{\ast}_{m'+1, {\bf R}}({\bf r})-\sqrt{m+1} \Psi_{m+1, {\bf R}}({\bf r}) \Psi^{\ast}_{m', {\bf R}}({\bf r})
\\
i\sqrt{m'+1} \Psi_{m, {\bf R}}({\bf r}) \Psi^{\ast}_{m'+1, {\bf R}}({\bf r})+i\sqrt{m+1} \Psi_{m+1, {\bf R}}({\bf r}) \Psi^{\ast}_{m', {\bf R}}({\bf r})
\end{array}
\right)
 \nonumber \\
\times
 \sum_{k=0}^{+ \infty} \frac{1}{k!} \,
\left(-\frac{l_{B}^2}{2} \Delta_{\bf R} \right)^{k}
g_{m ;m'}^{<}({\bf R},\omega)
\Bigg]
\label{simp}
\end{eqnarray}
\end{widetext}
where $
\rho({\bf r},\omega)=- i G^{<}({\bf r},{\bf r},\omega)
$
is the local spectral function.
Similar to Sec.~\ref{section3}, we wish to collect all contributions up to order
$l_B^2$ to the local current density.

\subsection{Electronic current at leading order}

The procedure to compute the different contributions to the current density is
completely analogous to the calculation of the electronic density done in
Sec.~\ref{section3}, although more lengthy. The leading contributions are
easily seen in Eq.~(\ref{simp}) to come from $g^{(0)}$ and $g^{(1)}$.

\subsubsection{Contribution from $g^{(0)}$: Density-gradient current}
Leading order Green's function is purely diagonal with respect to the Landau-level
index $m$, so that we have to consider combinations as
$\sqrt{m+1} \, \Psi_{m} \Psi_{m+1}^{\ast}$ in Eq. (\ref{simp}). Inserting
both $\rho^{(0)}$ from Eq.~(\ref{electrondensity}) and $g^{(0)}$ from Eq.~(\ref{g0}) (considering only the
contribution with $k=0$) and using the useful relation (\ref{practical}), we readily obtain
after the frequency integral
\begin{eqnarray}
{\bf j}^{(0)}({\bf r}) &= &
- \frac{e}{h} \hat{{\bf z}} \times {\bm \nabla}_{{\bf r}}
 \int\!\!\! \frac{d^{2}{\bf R}}{2 \pi l_{B}^{2}} \sum_{m=0}^{+\infty}
\hbar \omega_c n_{F}(\xi_{m}({\bf R})) \nonumber \\
&& \times
\left[
 \sum_{p=0}^{m}
\left| \Psi_{p,{\bf R}}({\bf r})
\right|^{2}
-
\frac{
\left| \Psi_{m,{\bf R}}({\bf r})
\right|^{2}}{2}
\right]
.
\label{j0gen}
\end{eqnarray}
This contribution to the current density has the property that its volume average
vanishes: $\int d^{2}{\bf r} \, {\bf j}^{(0)}({\bf r})={\bf 0}$.

As done previously for the local electronic density, the density-gradient
contribution~(\ref{j0gen}) can be expanded in the strict $l_B\rightarrow0$ limit
to recover a semiclassical expression,
\begin{equation}
{\bf j}^{(0)}({\bf r}) =
 \frac{e}{h}
 \sum_{m=0}^{+\infty}
\left(m+\frac{1}{2} \right) \hbar \omega_{c}
 {\bm \nabla}_{{\bf r}}
n_{F}(\xi_{m}({\bf r}))
 \times \hat{{\bf z}}
.
\label{j0}
\end{equation}
This result coincides with the formula for the ``edge'' electronic current
density derived within a different method in the Ref. \onlinecite{Geller1994}.
It has clearly the form of a current flow responding to a gradient of the
density. It thus vanishes in the incompressible regions where the density is
quasiconstant, and becomes important in the compressible regions of the system where
the local density is strongly inhomogeneous and the bare potential is almost perfectly
screened. Such regions are not necessarily located at the edges of the system,
but are rather spread throughout the system. The denomination of ``edge
current'' is thus in some sense abusive. 
Therefore, we prefer to call 
contributions~(\ref{j0gen}) and (\ref{j0}) a density-gradient current.

\subsubsection{Contribution from $g^{(1)}$: Drift current}
As emphasized in our previous paper \cite{Champel2007}, the well-known drift
contribution to the current density appears, in fact, beyond the limit $l_{B} \to
0$, i.e., when considering Green's functions $g^{(1)}$ which take into account the
first processes of Landau level mixing. Such a drift contribution is however of
the same order as the density-gradient contribution [the reason is that there
is a prefactor $l_{B}^{-1}$ in the general expression of the current
density, see the second term in the rhs of Eq. (\ref{simp})].

Using the general expression for the current density (\ref{simp}) with $k=0$
and inserting $g^{(1)}$ from Eq.~(\ref{g1}), we get
\begin{widetext}
\begin{eqnarray}
{\bf j}^{(1)}({\bf r}) &=&
\frac{e \hbar }{ 2 m^{\ast}} \hat{{\bf z}} \times {\bm \nabla}_{{\bf r}} \rho^{(1)}({\bf r})
+
\frac{e \hbar }{m^{\ast}}
\int\!\!\! \frac{d^{2}{\bf R}}{2 \pi l_{B}^{2}}
\sum_{m=0}^{+\infty}
\frac{
n_{F}(\xi_{m+1}({\bf R})) -n_{F}(\xi_{m}({\bf R}))}{\hbar \omega_{c}}
\left[
(m+1) \left| \Psi_{m+1,{\bf R}}({\bf r})\right|^{2}
 \hat{{\bf z}} \times {\bm \nabla}_{{\bf R}} V({\bf R})
 \right.
\nonumber
\\
&& \left.
+
\sqrt{m+1}\sqrt{m+2}
\left(
\begin{array}{c}
\mathrm{Im} \\
\mathrm{Re}
\end{array}
\right)
(\partial_{X} V+i \partial_{Y}V)
\Psi_{m,{\bf R}}({\bf r})
\Psi_{m+2,{\bf R}}^{\ast}({\bf r})
\right].
\label{j1}
\end{eqnarray}
\end{widetext}
In Appendix~\ref{appj} we provide a detailed calculation of this expression ~(\ref{j1}), which contains a peculiar
 term (the last one) involving vortex
wave functions with Landau indices that differ by 2. As a result, we find that the leading contribution to Eq. (\ref{j1})
reads

\begin{equation}
{\bf j}^{(1)}({\bf r}) =
\frac{e}{h} \!
\int\!\!\! d^{2}{\bf R} \!
\sum_{m=0}^{+\infty}
\left| \Psi_{m,{\bf R}}({\bf r})\right|^{2}
n_{F}(\xi_{m}({\bf R}))
 {\bm \nabla}_{{\bf R}} V({\bf R})
\times \hat{{\bf z}}
.
\label{j1fin}
\end{equation}
In the limit $l_{B} \to 0$, above contribution (\ref{j1fin}) yields
\begin{equation}
{\bf j}^{(1)}({\bf r})
=
\frac{e}{h}
\sum_{m=0}^{+\infty}
n_{F}(\xi_{m}({\bf r}))
 {\bm \nabla}_{{\bf r}} V({\bf r})
\times \hat{{\bf z}}
.
\label{drift}
\end{equation}
We thus recover the well-known drift current that can be found in the
literature~\cite{Geller1994}, while expression~(\ref{j1fin})
constitutes a quantum version of this drift current, that may be used at low
temperature.

\subsection{Electronic current at order $l_B^2$}

We aim at collecting exhaustively all contributions to the current density that
are proportional to $l_{B}^2$: This is the order where the first dissipative features
are expected to appear (see Sec.~\ref{section6}). We will simply list here the
various origins for these terms and refer the reader to Appendix~\ref{appj} for
the detailed calculation.

There is first a subdominant contribution coming from $g^{(1)}$, with expression ~(\ref{j1}).
In Appendix~\ref{appj} this contribution is denoted ${\bf j}^{(1)}_\mathrm{sub}$, and is given in
Eq.~(\ref{j1sub}). Another contribution of order $l_{B }^{2}$ arises with second-order 
Green's function $g^{(2)}$. The latter contains diagonal elements ($m=m'$) that combine with
the first and second terms in the rhs of Eq. (\ref{simp}), and also
off-diagonal elements $\delta_{m,m' \pm 2}$, which have to be inserted in the second
term of the rhs of Eq. (\ref{simp}). The final expression for
${\bf j}^{(2)}$, which also includes the contribution from the function $g^{(0)}$ appearing
with the term $k=1$ in Eq. (\ref{simp}), is given by Eq.~(\ref{j2nonlocal}).
Finally, the off-diagonal elements $\delta_{m,m' \pm 1}$ of $g^{(3)}$, calculated in
Appendix~\ref{apg3}, combine with the contribution from the function $g^{(1)}$
associated with the term $k=1$ in Eq. (\ref{simp}), giving the final
result for ${\bf j}^{(3)}$ in Eq.~(\ref{j3nonlocal}).

\subsection{Semiclassical current: The strict $l_B\rightarrow0$ expansion}
\label{semij}

As previously done with the local electronic density, it is also possible to express the current density
under the form of a strict expansion in powers of the magnetic length.
In this section, we want to obtain the corrections of order $l_{B}^{2}$ to the
well-known semiclassical expression~(\ref{drift}) for the drift current,
which is purely transverse.
All these subleading contributions are collected in Appendix~\ref{appj}. Since
the semiclassical expansion in $l_{B}$ is only valid in a ``high'' temperature
regime, we will only present here the terms proportional to the Fermi factor
that are dominant in this regime with respect to the other terms involving
derivatives of the Fermi factor.

Collecting Eq.~(\ref{drift}) with the contributions from
Eqs.~(\ref{j0local})-(\ref{j3local}), we get the
leading contribution to the semiclassical current,
\begin{widetext}
\begin{equation}
{\bf j}({\bf r}) =
\frac{e}{h} \sum_{m=0}^{+ \infty}
n_{F}(\xi_{m}({\bf r}))
\,
\left[
{\bm \nabla_{\bf r}} V +
l_{B}^{2}
\frac{ \left(
{\bm \nabla_{\bf r}} V \cdot
{\bm \nabla_{\bf r}}
\right)}{\hbar \omega_{c}}
{\bm \nabla_{\bf r}} V
+\frac{3}{2}l_B^2 \left(m+\frac{1}{2} \right) \Delta {\bm \nabla_{\bf r}} V
\right]
\times \hat{{\bf z}}.
\label{j2}
\end{equation}
\end{widetext}
This expression constitutes one of the main physical results of the paper and is further
analyzed in Sec.~\ref{section6} dealing with the nonequilibrium transport properties.

\section{Checking our theory: Comparisons with an exactly solvable case
\label{section5}
}

\subsection{One-dimensional parabolic confinement model}
The aim of this section is to benchmark our results for the local equilibrium charge
and current densities obtained with the vortex states. For this purpose a
comparison to the exact solution that can be obtained for
the case of a one-dimensional parabolic confining potential turns out to be
quite entlightening. We will therefore focus here on the following potential
profile:
\begin{equation}
V(x)=\frac{1}{2} m^{\ast} \omega_{0}^{2} x^{2}.
\end{equation}
Following Ref. \onlinecite{Geller1994}, the exact eigenstates and eigenenergies corresponding to
this particular choice of one-dimensional potential are given in the Landau
gauge ${\bf A}=B x \hat{{\bf y} }$ by
\begin{eqnarray}
\Psi_{n p}({\bf r})
&=&
\frac{e^{-ipy}\, e^{-\frac{\left(x-\frac{\omega_{c}}{\Omega} pL^{2}\right)^{2}}{2 L^{2}}} }{\sqrt{2^{n+1} n! \pi^{3/2} L}} \,
H_{n}\left(
\frac{x-\frac{\omega_{c}}{\Omega} pL^{2}}{L}
\right)
\label{hermite}
,\\
E_{np}&=&\hbar \Omega \left(n+\frac{1}{2}\right)+V(pL^{2}),
\end{eqnarray}
where $\Omega=\sqrt{\omega_{c}^{2}+\omega_{0}^{2}}$ and $L=\sqrt{\hbar/m^{\ast} \Omega}$
are the renormalized cyclotron pulsation and magnetic length respectively, and
$H_n$ denotes the $n^\mathrm{th}$ Hermite polynomial. These
wave functions, fully extended plane waves along constant energy contour while
strongly localized in the transverse $x$ direction, are certainly very different
from the vortex states [Eq.~(\ref{vortex})], which are localized in all directions
without preferred symmetry, so that this comparison provides a very stringent
test on the vortex theory.

Physical observables, such as the local electronic density, are readily obtained as
\begin{equation}
\rho(x) = \sum_{n=0}^{+ \infty} \int_{-\infty}^{+ \infty}\!\!\!\!\! dp \,
\left| \Psi_{n p}({\bf r})\right|^{2}
 n_{F}\left(E_{np}
\right),
\label{rhoexact}\\
\end{equation}
while the equilibrium current density, directed in the $y$ direction,
reads~\cite{Geller1994}:
\begin{eqnarray}
j(x)=\sum_{n=0}^{+ \infty}
\int_{- \infty}^{+ \infty}
dp \, j_{np} (x) n_{F}(E_{np}),
\end{eqnarray}
where
\begin{eqnarray}
j_{np}(x) & =& \frac{|e| \Omega^{2}}{\omega_{c}}
\left(
\frac{\omega_{c}}{\Omega}p L^{2}-x
\right)
\left| \Psi_{np}({\bf r})\right|^{2} \nonumber \\
&&
+ |e| \omega_{c} l_{B}^{2} \frac{V'(x)}{\hbar \omega_{c}}
\left| \Psi_{np}({\bf r})\right|^{2}
.
\label{jexact}
\end{eqnarray}

\subsection{Checking analytically the semiclassical expansion}

The first check, which is crucial for demonstrating the mathematical
consistency of our semiclassical limit, as obtained from the vortex
calculation, consists in developping both Eqs.~(\ref{rhoexact}) and (\ref{jexact})
in a strict magnetic length expansion at order $l_B^2$.
For this purpose, we develop $\Omega$ and $L$ using $\omega_0^2/\omega_c^2=l_B^2
V''(x)/\hbar\omega_c$, and perform a Taylor expansion. The resulting Gaussian integrals are computed using the formula
\begin{equation}
\int_{-\infty}^{+\infty} \!\!\!\!\! d \xi \, e^{-\xi^{2}} \xi^{2}
H_{n}^{2}(\xi)=\sqrt{\pi} \, 2 ^{n} n! \left( n+ \frac{1}{2}\right).
\end{equation}
This leads to the result
\begin{eqnarray}
\rho(x)& =&\frac{1}{2\pi l_B^2} \sum_{n=0}^{+\infty}
\left[
n_F(x) + l_B^2 n_F(x) \frac{V''(x)}{\hbar \omega_c}
\right.
\nonumber \\
&&
\left.
+ l_B^2 n_F'(x) V''(x)
+ \frac{l_B^2}{2} n_F'(x) \frac{(V'(x))^2}{\hbar\omega_c}
\right.
\nonumber \\
&&
\left.
+ \frac{l_B^2}{2} \left(n+\frac{1}{2}\right) n_F''(x) (V'(x))^2
\right],
\end{eqnarray}
where $n_{F}(x)=n_{F}\left[(n+1/2)\hbar \omega_{c}+V(x) \right]$,
which is obviously equivalent to Eq.~(\ref{densitysecondorder}).

The calculation of the semiclassical current density at order $l_B^2$ follows
the same lines, and using formula
\begin{equation}
\int_{-\infty}^{+\infty} \!\!\!\!\! d \xi \, e^{-\xi^{2}} \xi^{4}
H_{n}^{2}(\xi)=3 \sqrt{\pi} \, 2 ^{n-1} n! \left(n^{2}+n+1/2 \right),
\end{equation}
we recover the leading density-gradient and drift contributions,
\begin{eqnarray}
j_{0}(x)=
\frac{|e|}{h}
\sum_{n=0}^{+ \infty}
\left[
n_{F}(x) V'+\left(n+\frac{1}{2}\right) \hbar \omega_{c} n'_{F}(x) V'
\right],
\nonumber \\
\vspace*{-0.3cm}
\end{eqnarray}
while the terms of order $l_B^2$ read
\begin{widetext}
\begin{eqnarray}
j_{2}(x)=
\frac{|e|}{h} l_{B}^{2}
\sum_{n=0}^{+ \infty}
\left[
\frac{ n'''_{F}(x)}{4} \left(n^{2}+n+1/2 \right) \hbar \omega_{c} (V')^{3}
+ n''_{F}(x) \left( \frac{5}{4}n^{2}+\frac{5}{4}n + \frac{1}{2}\right) V' V'' \hbar \omega_{c}
+
n'_{F}(x)\frac{(V')^{3}}{2\hbar \omega_{c}}
\right. \nonumber
\\
+ \left.
 n''_{F}(x) \left( n + \frac{1}{2}\right) (V')^{3}
+
n'_{F}(x) \frac{7}{2}\left(n+\frac{1}{2}\right) V ' V''
+n_{F}(x) V' \frac{V''}{\hbar \omega_{c}}
\right].
\label{jsemi}
\end{eqnarray}
\end{widetext}
One can easily check that collecting all terms in
expressions~(\ref{j0local})-(\ref{j3local}) for a one-dimensional potential yields
the same result, giving a good confidence in the vortex method to generate the
semiclassical expansion. We emphasize however that our semiclassical results were
derived for an arbitrary two-dimensional potential, so that extra terms are
actually present in our semiclassical formula with respect to Eq.~(\ref{jsemi}).
In particular some of the terms appearing in Eq.~(\ref{j2}) involves derivatives
of the potential in two orthogonal directions and cannot be infered from this
simple calculation of a one-dimensional parabolic potential.
It is interesting to note that in nonequilibrium it is precisely these additional terms in the current density that are seemingly associated with dissipative features (see Sec. \ref{section6}).

\subsection{Systematic numerical comparison to the vortex theory}
\label{comp}

We aim here at giving a more quantitative comparison for the electronic density
between the exactly solvable model and the various expansions discussed in
Sec.~\ref{section3}. We therefore compute numerically the
expression~(\ref{rhoexact}) and investigate both the semiclassical
approximation~(\ref{electrondensity2}) and the order $l_B^2$ quantum
expressions~(\ref{electrondensity}) and (\ref{rho2diag}). Note that the
$l_B^2$ corrections [Eq.~(\ref{densitysecondorder})] to the semiclassical result
present derivatives of the Fermi factor, which are either very small (at high
temperatures) or very singular at low temperatures, so that they are not included 
in the comparison.
In contrast, the $l_B^2$ corrections [Eq.~(\ref{rho2diag})] to the leading quantum
result~(\ref{electrondensity}) are more regular, and their inclusion is
important to reach quantitative agreement in an intermediate temperature range, as 
we will demonstrate below.

In order to show that the reliability of these different approximation schemes are rooted
in specific temperature regimes, we present results for different temperatures at a given
confinement energy $\hbar\omega_0=\omega_c/5$, small enough to ensure the smoothness of the
external potential, yet already sufficiently large so that the semiclassical
approximation is in trouble at low temperature. The electrochemical potential is
also fixed by taking $\mu^\ast=3\omega_c$, so that three Landau levels are present at 
the center of the system.

\begin{figure}[ht!]
\includegraphics[width=8.5cm]{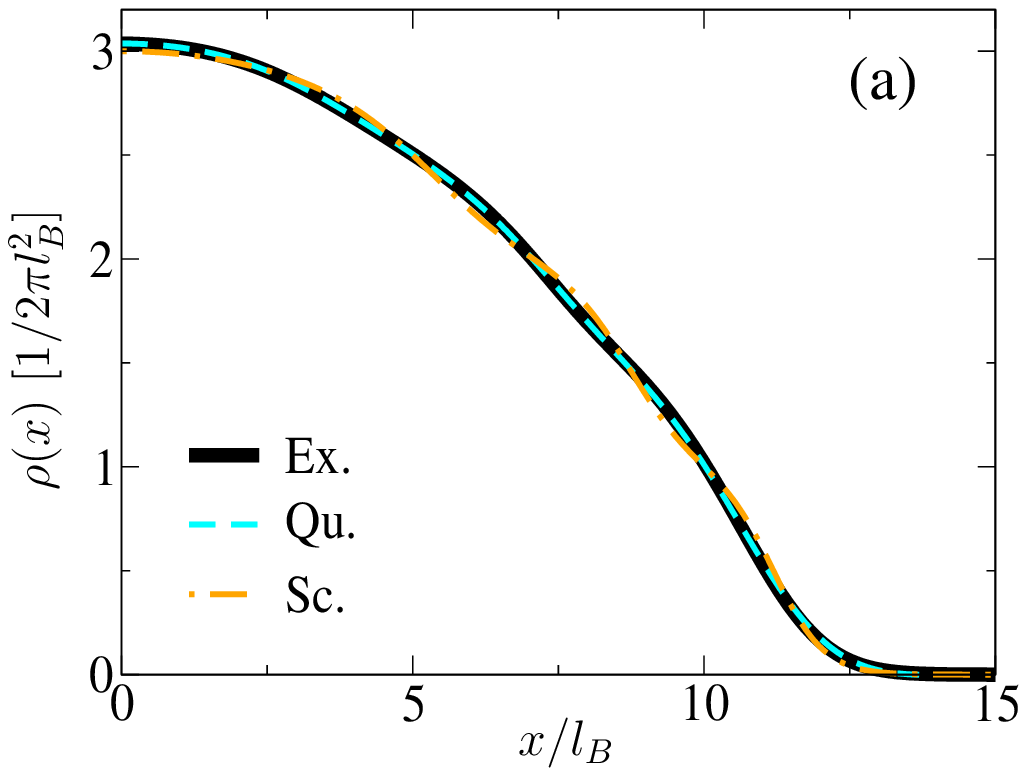}
\includegraphics[width=8.5cm]{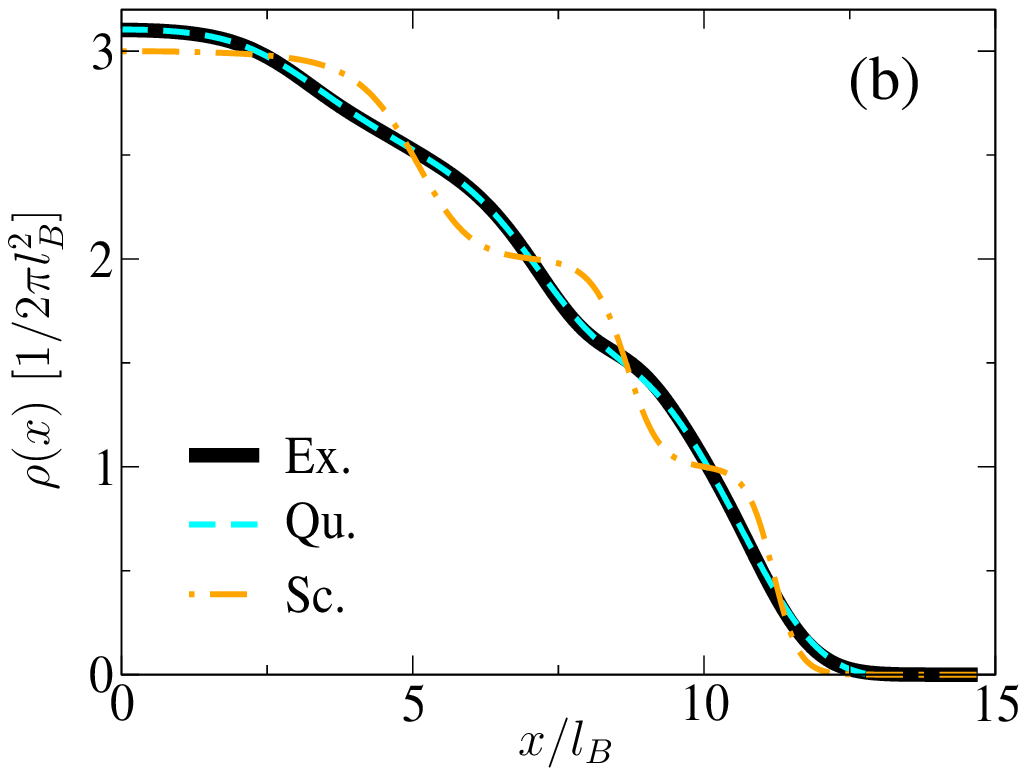}
\includegraphics[width=8.5cm]{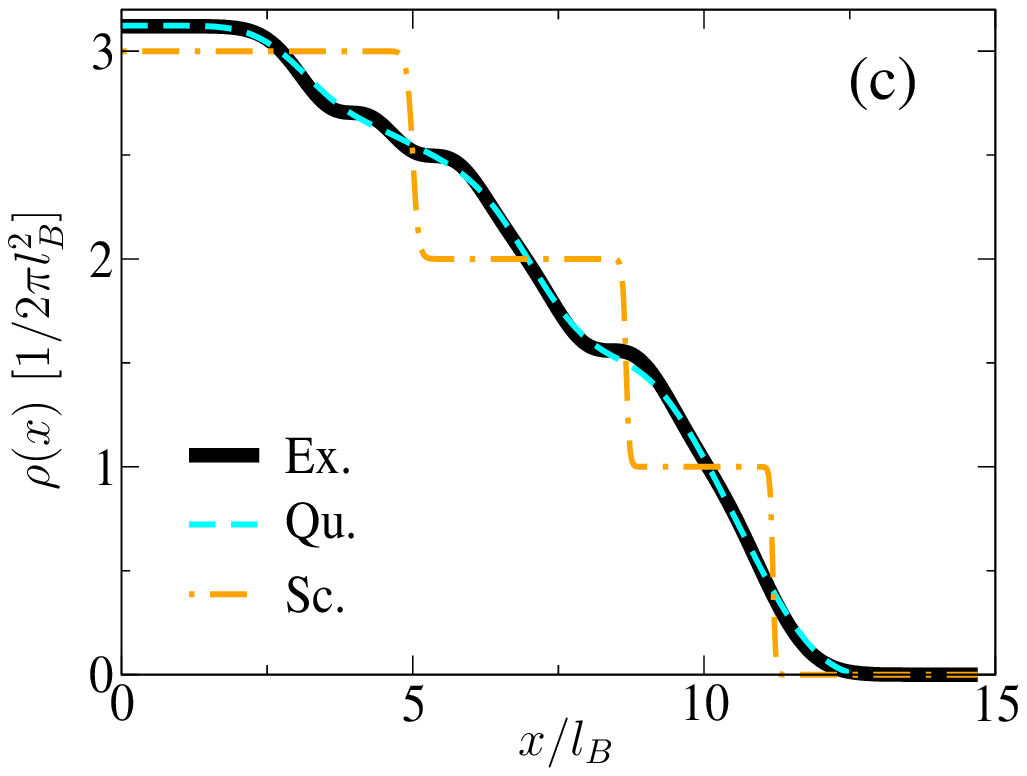}
\caption{ (Color online) Local electronic density $\rho(x)$ in units of
$1/2\pi l_{B}^{2}$
for the one-dimensional parabolic potential with $\omega_{0}=\omega_c/5$ and
$\mu^\ast=3\omega_c$, as a function of $x/l_B$, comparing exact
expression~(\ref{rhoexact})
(solid curve, label Ex.) with the semiclassical
expansion~(\ref{electrondensity2}) (dotted-dashed curved,
label Sc.) and the quantum expansion given by Eqs. ~(\ref{electrondensity}), (\ref{rho1eq}) and
(\ref{rho2diag}) (dashed curve, label Qu.).
The three different panels (a)-(c) correspond to decreasing
temperatures
$k_B T/\hbar\omega_c=0.2,0.1$, and $0.01$.
}
\label{comparison}
\end{figure}

At temperatures not too low compared to the cyclotron frequency (first panel (a) of
Fig.~\ref{comparison} for $k_B T=0.2 \hbar \omega_c$), the semiclassical result is
still close to the exact solution, exactly matched by the quantum result. Lowering 
further the temperature (second panel (b) of Fig. \ref{comparison} for $k_B T=0.1 \hbar \omega_c$) 
shows increasing deviations with the semiclassical result, while the complicated
variations in the exact density are perfectly reproduced by the quantum formula.
In particular, both the small compressibility in the filling factor $n=3$ plateau [the density of the third filled Landau level is slightly greater than the value $ 3 \times (2 \pi l_{B}^{2})^{-1}$]
and the broad smearing of the $n=1$ and $n=2$ plateaus are quantitatively
described. In the very low temperature regime, small shoulders appear at
fractional densities (panel (c) of Fig. \ref{comparison} for
$k_B T=0.01 \hbar \omega_c$), which are associated with the zeros of the Hermite
polynomials in Eq.~(\ref{hermite}). These variations are only partially reproduced
by the quantum expression, but the overall agreement remains very good.

\subsection{Zero temperature limit: Resummation of the quantum development}
\label{resum}

We finally motivate the need for a resummation of the quantum expression to arbitrary
order in $l_B$ in the very low temperature regime, as hinted in
Sec.~\ref{remarks}.
As the semiclassical expression for the electronic density~(\ref{rhofinal}) is
clearly divergent at low temperature, one can indeed ask whether the leading quantum
result~(\ref{electrondensity}) and its order $l_B^2$ corrections [Eqs. (\ref{rho1eq})
and (\ref{rho2diag})] give satisfactory results for all temperatures.
Despite the excellent agreement observed above, the Fermi factor derivatives
appearing in these order $l_B^2$ terms tend to give important and uncontrolled
contributions in the zero-temperature limit.
To see this, let us forget for the time being the (negligible) terms inversely proportional
to $\hbar \omega_c$ in the quantum expression for the density, which then simply
reads
\begin{eqnarray}
\rho({\bf r}) & =&
\int\!\!\! \frac{d^{2}{\bf R}}{2 \pi l_{B}^{2}}
 \sum_{m=0}^{+ \infty}
\left|
\Psi_{m,{\bf R}}({\bf r})
\right|^{2}
\left[
n_{F}\left( \xi_{m}({\bf R})\right) \frac{}{}
\right.
\nonumber \\
&&
\left.
-
\frac{l_{B}^{2}}{4} \Delta_{\bf R}
n_{F}\left( \xi_{m}({\bf R})\right)
\right] \label{corrterm0} \\
&=& \rho^{(0)}({\bf r})-\frac{l_B^2}{4}\Delta_{\bf r}\rho^{(0)}({\bf r}).
\label{corrterm}
\end{eqnarray}
Here we have made an integration by parts to rewrite the second term in the
rhs of Eq. (\ref{corrterm0}). Because $\rho^{(0)}$ cannot change on a scale
smaller than $l_B$, as is clear from Eq. ~(\ref{electrondensity}), the $l_{B}^{2}$
correcting term in the rhs of Eq.~(\ref{corrterm}) cannot become singular in the
zero-temperature limit, in contrast to the semiclassical
expression~(\ref{rhofinal}). However, $\rho^{(0)}$ does change on the scale
$l_B$ at the boundary of an incompressible region at very low temperature, so that the correction
becomes of order one and needs to be resummed to all orders. The need for a
resummation is mathematically related to the fact that nonlocal vortex Green's
function has been developped at coinciding points in Eq.~(\ref{devg}), while keeping
a finite number of contributions. A clear example of such a nonlocal resummation
to all orders is the relation~(\ref{electronGreensimp}) between vortex and electron
propagators. In fact, the correction in Eq.~(\ref{corrterm}) is the combination of
the $-(l_B^2/2) \Delta_{\bf R} g^{(0)}$ term in Eq.~(\ref{electronGreensimp}) and the
$(l_B^2/4) \Delta_{\bf R} g^{(0)}$ contribution that can be extracted from $g^{(2)}$
in Eq.~(\ref{g2}).

\begin{figure}[bht!]
\includegraphics[width=8.5cm]{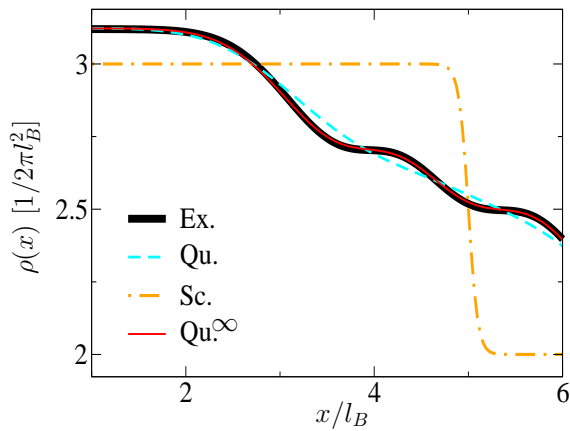}
\caption{ (Color online) Local electronic density $\rho(x)$ for the same
parameters as in Fig.~\ref{comparison}(c), focusing on the compressible region between filling factors $n=2$ and $n=3$.
This compares exact expression~(\ref{rhoexact}) (thick solid curve, label
Ex.),
undistinguishable from quantum expression (96) resummed to infinite order in $l_{B}$
(thin solid curve, label Qu.\hspace{-0.1cm}$^\infty$), with 
semiclassical expansion~(\ref{electrondensity2}) (dotted-dashed curve,
label Sc.) and
the quantum expansion up to order $l_{B}^{2}$ given by Eqs. ~(\ref{electrondensity}), (\ref{rho1eq}) and (\ref{rho2diag})
(dashed curve, label Qu.).
}
\label{fig2}
\end{figure}

By inspecting the recursion relation~(\ref{recursive}) in the small $l_B$ limit,
it is possible to infer that this class of most singular terms in the vortex propagator
is given {\it to all orders} by
\begin{equation}
g^{B\rightarrow\infty}_{m;m'}({\bf R})
= \delta_{m,m'} \sum_{k=0}^{+\infty} \frac{1}{k!} \left(\frac{l_B^{2}}{4}
\Delta_{\bf R}\right)^k
g^{(0)}_{m;m}({\bf R})
\label{gresummed}
\end{equation}
so that their combination with Eq.~(\ref{electronGreensimp}) leads to the final
quantum expression for the density in the small but nonzero $l_B$ limit,
\begin{eqnarray}
\rho^{B\rightarrow\infty}({\bf r}) &=&
\int\!\!\! \frac{d^{2}{\bf R}}{2 \pi l_{B}^{2}}
 \sum_{m=0}^{+ \infty}
\left|
\Psi_{m,{\bf R}}({\bf r})
\right|^{2}
\nonumber \\
&& \times
\sum_{k=0}^{+\infty} \frac{1}{k!} \left(-\frac{l_B^{2}}{4}
\Delta_{\bf R}\right)^k
n_{F}\left( \xi_{m}({\bf R})\right). \hspace*{0.5cm}
\end{eqnarray}
Using an integration by parts, this equation can be written in the equivalent form
\begin{eqnarray}
\rho^{B\rightarrow\infty}({\bf r}) &=&
\int\!\!\! \frac{d^{2}{\bf R}}{2 \pi l_{B}^{2}}
 \sum_{m=0}^{+ \infty}
n_{F}\left( \xi_{m}({\bf R})\right)
\nonumber \\
&& \times
\sum_{k=0}^{+\infty} \frac{1}{k!} \left(-\frac{l_B^{2}}{4}
\Delta_{\bf R}\right)^k
\left|
\Psi_{m,{\bf R}}({\bf r})
\right|^{2}
. \hspace*{0.5cm}
\end{eqnarray}
We note from Fourier analysis that the differential operator 
$$
\sum_{k=0}^{+\infty} \frac{1}{k!} \left(-\frac{l_B^{2}}{4}
\Delta_{\bf R}\right)^k
$$
 is nothing else than a
convolution operator
with the kernel $e^{-{\bf u}^2/4t}/(4\pi t)$, where $t=-l_B^2/4$. We can apply this
to the vortex density, to find
\begin{eqnarray}
\rho^{B\rightarrow\infty}({\bf r}) & = &
\int\!\!\! \frac{d^{2}{\bf R}}{2 \pi l_{B}^{2}}
 \sum_{m=0}^{+ \infty}
n_{F}\left( \xi_{m}({\bf R})\right)
\left| \Phi_{m,{\bf R}}({\bf r}) \right|^{2}
 \hspace*{0.5cm}
\label{rhoinfty2}\\
\left| \Phi_{m,{\bf R}}({\bf r}) \right|^{2} & = &
\int\!\!\! \frac{d^{2}{\bf u}}{4 \pi t} e^{-{\bf u}^2/4t}
\left| \Psi_{m,{\bf R-u}}({\bf r}) \right|^{2}
.
\label{Phiwave}
\end{eqnarray}
Performing formally the remaining Gaussian integral over ${\bf u}$ in Eq. (\ref{Phiwave}), we find
\begin{eqnarray}
\rho^{B\rightarrow\infty}({\bf r}) &=&
\int\!\!\! \frac{d^{2}{\bf R}}{2 \pi l_{B}^{2}}
 \sum_{m=0}^{+ \infty}
\frac{ n_{F}\left( \xi_{m}({\bf R})\right)}{\pi m!l_{B}^2}
A_m({\bf R-r}) \nonumber \\
&& \times
\exp\left[ -\frac{({\bf R}-{\bf r})^2}{ l_{B}^{2}} \right],
 \hspace*{0.5cm}
\label{rhoinfty}
\end{eqnarray}
where $A_m$ is the following polynomial:
\begin{eqnarray}
A_m({\bf R}) = \frac{\partial^m}{\partial s^m}
\left( \frac{1}{1+s}
\exp\left[ \frac{{\bf R}^2}{ l_{B}^{2}}\frac{2s}{1+s} \right] \right)_{s=0}
.
\end{eqnarray}
Final expression~(\ref{rhoinfty}) for the density is applicable down to zero
temperature, and provides the leading contribution in the small $l_B$ limit. In the case of a
one-dimensional potential $V(x)$, it is easy to check that the Gaussian integral
over the coordinate $Y$ in Eq.~(\ref{rhoinfty}) leads to the expected Hermite polynomials.
Regarding the remaining contributions that can be gathered from Eqs.~(\ref{rho1eq})
and (\ref{rho2diag}), involving Landau level mixing processes, a complete resummation scheme
amounts to extra shifts in the energies, as discussed in Sec.~\ref{semirho}. A final comparison is
given in Fig.~\ref{fig2}, which shows that, as long as $\omega_0\ll\omega_c$, 
these improved quantum expressions are undistinguishable from the exact result.

\section{Nonequilibrium properties
\label{section6}
}

\subsection{Distribution function and irreversibility}

We have solved so far Hamiltonian (\ref{Hamiltonian}) within the high magnetic field expansion without fully specifying the potential-energy term $V({\bf r})$.
This scheme allows us to study the equilibrium and nonequilibrium
situations on an equal footing. At equilibrium, $V({\bf r})$ consists of a fixed
background potential (including a confinement potential and an impurity random
potential) and of a Hartree potential  resulting
from the mutual Coulomb interactions between the electrons. As a result of a self-consistent calculation, this yields 
a  global effective electrostatic potential $V_{\mathrm{eff}}$ associated with
local microscopic electric fields. In the
nonequilibrium case, there is in addition an external potential-energy contribution
reflecting the appearance of macroscopic electric fields and of macroscopic chemical-potential gradients in the system induced by the presence of a macroscopic current flow. 
Within the nonequilibrium regime, which is considered from now on in this section, the
potential term $V({\bf r})$ in the Hamiltonian (\ref{Hamiltonian}) consists
thus of two different parts,
\begin{equation}
V({\bf r})=V_{\mathrm{eff}}({\bf r})+e \Phi({\bf r})
.
\label{Phi}
\end{equation}
Here  $\Phi$ is the nonequilibrium electrochemical potential that now varies in space. The latter
term takes into account the presence of a macroscopic electromotive field
${\bf E} =- {\bm \nabla_{\bf r}} \Phi$.

In Ref. \onlinecite{Champel2007} we
have solved the equation of motion for  correlation Green's function $G^{<}$ in
the vortex representation using the high magnetic field expansion, and have established that latter Green's
function expressed in the vortex variables is related to retarded and
advanced Green's functions at any order of our expansion in the
nonequilibrium stationary regime as
\begin{equation}
- i G^{<}_{\nu_{1};\nu_{2}}(\omega)=i
n_{F}(\omega)\left(G^{R}_{\nu_{1};\nu_{2}}(\omega)-G^{A}_{\nu_{1};\nu_{2}}(\omega)
\right).
\label{corr}
\end{equation}
In the electronic representation, the quantity $-i G^{<}$ which has the
character of a distribution function thus becomes

\begin{eqnarray}
-i G^{<}({\bf r},{\bf r}',\omega) &=
& \sum_{\nu_{1}, \nu_{2}}
\Psi_{\nu_{2}}^{\ast}({\bf r}') \Psi_{\nu_{1}}({\bf r})
\, i n_{F}(\omega)
\nonumber \\
&& \times
\left(G^{R}_{\nu_{1};\nu_{2}}(\omega)-G^{A}_{\nu_{1};\nu_{2}}(\omega)
\right)
 \label{electronGreenlesser2}
\end{eqnarray}
where we have used Eq. (\ref{corr}). The fact that the same relation
[Eq. (\ref{electronGreenlesser2})] holds in high magnetic fields in the equilibrium
regime as well as in the nonequilibrium stationary regime can be understood as
the realization of a local hydrodynamic equilibrium (or quasiequilibrium). This
result, which has been established from the microscopic derivation of the
quantum kinetic equation (see Ref. \onlinecite{Champel2007}), is physically expected
given that the microscopic characteristic lengthscale for the electron gas,
namely, $l_{B} \propto B^{-1/2} $, becomes in high magnetic fields the shortest
lengthscale. This means that it is possible to divide the system within a
continuum description into elementary subsystems which are almost isolated from
each other, permitting the introduction of thermodynamic variables depending on
the space variable ${\bf r}$ (see, e.g., Ref. \onlinecite{Akera2005}  and references therein).

It is worth mentioning that we have not introduced so far in the resolution of
the Dyson equation any averaging to account for the presence of a random
potential. In fact, this is not needed at this level since within the
high magnetic field expansion all the physics in the vortex representation
appears to be purely local, the Hamiltonian being diagonalized in a closed-form
order by order in powers of the magnetic length with the use of local vortex
Green's functions $g_{m_{1};m_{2}}({\bf R})$. Dissipation and irreversibility,
that are usually introduced already at the level of the Dyson equation with the
impurity averaging procedure to account for the presence of random scattering
interactions in zero or weak magnetic fields, take its roots within a different
mechanism in high magnetic fields.

In fact, the stochastic character 
is intrinsic to our high magnetic field expansion making use of the vortex basis, and we can associate somehow the transformation from the vortex
to the electronic representations with a loss of information (thus irreversibility) provided that there exists some dynamical instability in the system. 
Indeed, by
solving the Dyson equation in the vortex representation, we have basically
augmented the set of allowed quantum numbers since the vortex basis is
overcomplete. Note however that the expansion of the matrix elements of the
potential in the vortex representation is granted in high fields due its
unicity, which clearly results from the possibility to truncate the series
expansion in the magnetic length $l_{B}$, see Ref. \onlinecite{Champel2007}. Coming back to the electron
representation, an indeterminacy 
illustrated
by the presence of weight factors, namely, the wave functions $\Psi_{\nu}({\bf
r})$ in Eq. (\ref{electronGreenlesser2}), appears, giving rise to a
statistical-like description.
 Our high magnetic field expansion alone contains
thus microscopically the stochastic character which is a prerequisite 
ingredient for the expression of a loss of information. This loss becomes effective as soon as a dynamical instability associated with a divergence of neighboring trajectories exists in the system. This general view is corroborated by explicit calculations in the following. In Sec. \ref{Space}, we shall derive a microscopic expression for the conductivity tensor that is indeed
associated in the nonequilibrium regime  with an electrochemical potential drop 
occuring only in the vicinity of a local instability of the dynamics which, in the present case, is brought by the presence of saddle-points of the local equilibrium density.

It is worth noting finally that we do not take into account 
 the interaction of the system with an
external environment. This alternative approach to dissipation 
considering the system plus reservoir  couples the relevant quantum system to a large
number of environmental degrees of freedom, such as phonons or quantum fluctuations of the electromagnetic field. Dissipation arises then because the system of interest can exchange energy with the rest of the larger system. More precisely, a loss of information is usually explicitly introduced in the calculations when tracing out these environmental degrees of freedom. We do not consider this mechanism of dissipation as
being the most relevant here.
 On the contrary, the  irreversibility mechanism described in this paper takes place in the bulk of the system iself, i.e. in the two-dimensional electron gas. This  is a fundamental and key point
in our transport theory. The relevance of a given  approach to dissipation can finally be appreciated at the level of the comparison between theory and experiments, since the dissipative transport properties are strongly and
intrinsically related to its dissipation mechanisms.

\subsection{Nonequilibrium current density}

 Since
the two potential-energy terms $V_{\mathrm{eff}}$ and $e \Phi$ in Eq. (\ref{Phi}) can be treated
technically on an equal footing in our high magnetic field theory, the
nonequilibrium current $\delta \, {\bf j}$ can, in fact, be rather
straightforwardly deduced from the expressions of the equilibrium current
density derived in the former Sec. \ref{section4} and Appendix \ref{appj}.

The primary goal of this paper is not to provide a full quantitative analysis of
macroscopic transport properties but just to show that our theory does contain
information on microscopic dissipative mechanisms, and thus allows us to fully
determine the spatial dependence of the electrochemical potential $\Phi({\bf
r})$. For the sake of simplicity, we shall therefore restrict ourselves to the
regime where the current density can be expressed in a local form [local
expressions (\ref{j0}), (\ref{drift}) and (\ref{j0local})-(\ref{j3local})]. This regime does not correspond to
the lowest temperatures for which the nonlocal nature of the current density
associated with quantum tunneling becomes predominant.

It is clear, e.g., from formulas (\ref{j0}) and (\ref{drift}), that a nonequilibrium current
can be generated in the linear response and at a uniform temperature by
simultaneous density and electrostatic potential variations. This indicates that in principle
we can separate the total current into two different contributions. One
contribution corresponds to the diffusion current (terms involving gradients of
the density) whose physical origin is associated with the tendency of the system
to make the density uniform. The other contribution represents the current
produced by the electric field which accelerates the electrons: It corresponds to electrical conduction which
occurs by definition for a uniform density. The true driving force for the
electrons is finally a combination of chemical and electrostatic potentials differences,
i.e., it is characterized by the variations of the electrochemical potential
$\Phi$ which yield at the macroscopic scale a voltage drop. Although the roles of the electrostatic and chemical potentials are fundamentally different microscopically, the precise composition of $\Phi$ is irrelevant in the linear response regime. Indeed, the integrated nonequilibrium  electrical current can be seen, e.g.,  either as resulting entirely from a density-gradient current, or equivalently, as being entirely produced by macroscopic electrostatic variations (note that this equivalence is only valid in the linear response).
In this paper, we shall adopt the point of view of the conduction mechanism, i.e., the case where a transport current $\delta \, {\bf j}$
is only sustained by electrostatic variations (the electrochemical potential
changes are only identified with the electrostatic potential changes in the system).

Any analysis (semiclassical or quantum) of the nonequilibrium properties where
interaction effects are expected to play a crucial role involves the
simultaneous resolution of a transport equation and of the Poisson equation (in
this direction, see, e.g., Ref. \onlinecite{Guven2003}). Since the equilibrium and
nonequilibrum regimes can be described with almost the same expressions, we can
first consider at a qualitative level that the interaction effects in the
nonequilibrium case do not differ substantially from that known in the
equilibrium case. The screening in the compressible regions being almost
perfect at low temperature, we expect a priori that the nonequilibrium conduction current is  principally confined to the incompressible regions where most of the macroscopic electrostatic variations giving rise to
voltage drops can occur (in other terms, this means that only one type of currents - diffusion or conduction - contributes to a given region in the ideal case of perfect compressibility and incompressibility). This aspect
concerning the nonequilibrium conduction current distribution has already been put forward
by different authors \cite{Tsemekhman1997,Guven2003,Siddiki2004}.

Using Eqs. (\ref{Phi}), (\ref{j0}), (\ref{drift}) and (\ref{j0local})-(\ref{j3local}), and keeping only the terms that
are linear in variations of the electrochemical potential $\Phi$ (since we
consider the linear response) and that do not contain derivatives of the Fermi
function factor (we consider the conduction mechanism which involves the whole Fermi sea), we get at leading
order for the nonequilibrium conduction current

\begin{eqnarray}
\delta \, {\bf j}_{0}({\bf r})
=
\frac{e^{2}}{h} \sum_{m=0}^{+ \infty}
n_{F}(\xi_{m}({\bf r}))
\,
{\bm \nabla}_{\bf r} \Phi
\times \hat{{\bf z}}.
\label{leading}
\end{eqnarray}
From Eq. (\ref{j2}), we get a correcting contribution to
the nonequilibrium current $\delta {\bf j}$ which is second order in $l_{B}$,
\begin{widetext}
\begin{eqnarray}
\delta \, {\bf j}_{2}({\bf r})
=
\frac{e^{2}}{h} \sum_{m=0}^{+ \infty}
n_{F}(\xi_{m}({\bf r}))
\,
l_{B}^{2}
\left[
\frac{ \left(
{\bm \nabla_{\bf r}} \Phi \cdot
{\bm \nabla_{\bf r}}
\right)}{\hbar \omega_{c}}
{\bm \nabla_{\bf r}} V_{\mathrm{eff}}
+
\frac{ \left(
{\bm \nabla_{\bf r}} V_{\mathrm{eff}} \cdot
{\bm \nabla_{\bf r}}
\right)}{\hbar \omega_{c}}
{\bm \nabla_{\bf r}} \Phi
+\frac{3}{2}\left(m+\frac{1}{2} \right) \Delta_{{\bf r}} {\bm \nabla_{\bf r}} \Phi
\right]
\times \hat{{\bf z}}.
\label{transcorr}
\end{eqnarray}
\end{widetext}
From now on the Fermi factor $n_{F}(\xi_{m}({\bf r}))$ is a functional of the
effective equilibrium potential $V_{\mathrm{eff}}$ which differs only slightly
from the bare potential in the incompressible regions of the
system where the screening is ineffective. A smooth spatial variation of the
factor  $n_{F}(\xi_{m}({\bf r}))$ exists in these regions as a result of the finite temperature.

Obviously, the leading contribution (\ref{leading}) yields local Ohm's law
which takes the form
\begin{equation}
\delta \, {\bf j}_{0}({\bf r})
= \hat{\sigma}({\bf r}) \, {\bf E}({\bf r})
=
\sigma_{H}({\bf r}) \, \hat{{\bf z}} \times {\bf E}({\bf r}),
\label{trans0}
\end{equation}
with a local conductivity tensor containing only the transverse Hall component
\begin{equation}
\sigma_{H}({\bf r})
=
\frac{e^{2}}{h}
\sum_{m=0}^{+ \infty}
n_{F}\left(\xi_{m}({\bf r}) \right),
\label{Hall}
\end{equation}
where $\xi_{m}({\bf r})=E_{m}+V_{\mathrm{eff}}({\bf r})$.
Ohm's law (\ref{trans0})  with local Hall coefficient (\ref{Hall}) is already well-known and is
used in most of the existing transport theories discussing the integer quantum
Hall effect. The absence of diagonal components for the conductivity tensor in this (semiclassical) limit
$l_{B} \to 0$ is rather welcome, since it is compatible with an extremely small
longitudinal resistance as observed when the Hall resistance presents plateaus.
However, this absence points out at the same time an insufficiency of the
formula (\ref{trans0}) to describe the transition region between the Hall
plateaus when high peaks of the longitudinal (dissipative) magnetoresistance are
seen. This insufficiency is, in fact, cured when considering contribution (\ref{transcorr}) to the current arising from the next order terms in the $l_{B}$ expansion, as shown further. At a general level, we note that our quantum-mechanical derivation of the transport current
justifies on a microscopic basis the use of phenomenological models assuming a
local conductivity tensor
\cite{Ruzin1993,Cooper1993,Dykhne1994,Simon1994,Ruzin1995,Guven2003,Siddiki2004,Ilan2006}
that have been considered so far to explain successfully some transport features
of the quantum Hall effect.

To our knowledge the first quantum corrections [Eq. (\ref{transcorr})] to  Ohm's
law (\ref{trans0}) had not been derived before in the literature. We find that
they contain local corrections (which give rise to both transverse and diagonal
components in the local conductivity tensor) as well as nonlocal corrections
(terms involving second- and third-order derivatives of $\Phi$). These nonlocal
terms can be viewed as fingerprints of the nonlocal
quantum tunneling processes in the considered semiclassical regime.

\subsection{Spatial dependence of the electrochemical potential
\label{Space}
}

The expansion of the current density in powers of $l_{B}$ has led us quite
naturally to a local continuum description of current conduction. Within this
``classical'' picture of transport (our theory is nevertheless developed in a fully quantum
mechanical framework), the stationary equation of continuity ensuring the charge
conservation
\begin{equation}
{\bm \nabla_{\bf r}} \cdot {\bf j} =0,
\label{continuity}
\end{equation}
supplemented by boundary conditions, constrains the spatial dependence of the
electrochemical potential when applied to the nonequilibrium current density
$\delta \, {\bf j}$ provided that some dissipation mechanisms are accounted
for within the considered expressions for $\delta \, {\bf j}$ [note that Eq. (\ref{continuity}) becomes an identity for the equilibrium current
density, as can be easily checked].

Inserting the leading contribution $\delta \, {\bf j}_{0}$ [Eq. (\ref{trans0})]
into Eq. (\ref{continuity}) and using ${\bf E}=- {\bm \nabla_{\bf r}} \Phi$, we get the
equation
\begin{eqnarray}
\left( {\bm \nabla_{\bf r}} \sigma_{H} \times {\bm \nabla_{\bf r}} \Phi \right) \cdot \hat{{\bf z}}
= 0,
\label{nice}
\end{eqnarray}
which has been thoroughly discussed in the literature
\cite{Ruzin1993,Cooper1993,Dykhne1994}. From this equation, it turns out that
the electrochemical potential lines and the lines of constant $\sigma_{H}$ must
coincide. It is worth noting that the condition (\ref{nice}) is automatically
obeyed at the critical points of $\sigma_{H}$ which correspond also to the
critical points of the density, or of the potential $V_{\mathrm{eff}}$. This
means that there still exists a degeneracy in the vicinity of these critical
points, which has to be lifted. Although being small, the correcting
contributions [Eq. (\ref{transcorr})] will play this important role of dictating locally the
spatial dependence of $\Phi({\bf r})$, as we prove now.

Since we are considering the nonequilibrium current density in the neighborhood
of ${\bm \nabla_{\bf r}} V_{\mathrm{eff}} = 0 $, we can first safely ignore in
Eq. (\ref{transcorr}) the term proportional to ${\bm \nabla_{\bf r}}
V_{\mathrm{eff}}$ and which involves the second-order derivative of $\Phi$. At a
preliminary stage, we shall also disregard the other nonlocal term (with the
third-order derivative of $\Phi$), and justify this assumption a posteriori.
Consequently, the second-order contribution to the current reduces to

\begin{eqnarray}
\delta \, {\bf j}_{2}({\bf r})
\approx
\frac{e^{2}}{h} \sum_{m=0}^{+ \infty}
n_{F}(\xi_{m}({\bf r}))
\,
l_{B}^{2}
\frac{ \left(
{\bm \nabla_{\bf r}} \Phi \cdot
{\bm \nabla_{\bf r}}
\right)}{\hbar \omega_{c}}
{\bm \nabla_{\bf r}} V_{\mathrm{eff}}
\times \hat{{\bf z}}.
\nonumber \\
\vspace*{-0.3cm}
\label{transcorr2}
\end{eqnarray}
Combining this Eq. (\ref{transcorr2}) with Eq. (\ref{trans0}), we get local Ohm's law
with a local conductivity tensor being given by
\begin{equation}
\hat{ \sigma}({\bf r})= \sigma_{H}({\bf r}) \left ( \begin{array}{cc}0 & -1 \\
1 & 0
 \end{array}\right)
\left(
 \hat{ 1}
+\frac{l_{B}^{2}}{\hbar \omega_{c}}
\hat{{\cal H}} \left[V_{\mathrm{eff}} \right]({\bf r})
\right),
\label{ohm}
\end{equation}
where $\hat{1}$ is the 2 x 2 identity matrix and $\hat{{\cal H}}$ is
the Hessian matrix of the function $V_{\mathrm{eff}}$, i.e., we have
$\hat{{\cal H}} \left[V_{\mathrm{eff}} \right]_{ij}({\bf r})
=\partial_{i j}^{2} V_{\mathrm{eff}}({\bf r})$.

We remark that the local conductivity tensor does not exhibit the usual
symmetries, i.e., the Onsager-Casimir reciprocity relations. For example, we find
here that generally $\sigma_{xx}({\bf r})=-\sigma_{yy}({\bf r})$ and $\sigma_{xy}({\bf r}) \neq - \sigma_{yx}({\bf r})$, whereas the Onsager relations
imply $\sigma_{xx}=\sigma_{yy}$ and $\sigma_{xy}=-\sigma_{yx}$. In fact, it is worth noting that the Onsager
relations result from fingerprints of the time-reversal invariance of the
microscopic equations after some averaging procedure (see, e.g., Ref. \onlinecite{Akera2005}). In the present case, we have
derived a local conductivity tensor from the microscopic equations without
resorting to any averaging procedure. Obviously, the local terms involving the Hessian matrix contribution vanish in the volume average; the Onsager relations are then restored. This indicates that the Hessian matrix terms can be interpreted as a result of {\em local fluctuations} only. Let us  also emphasize that current (\ref{transcorr2}) purely stems from Landau level mixing processes.
The found sign difference between the two
local diagonal components which appears as rather unexpected and unconventional
could be seen as a reminiscence of the antisymmetry imposed by the Lorentz
force, antisymmetry which is usually only exhibited by the Hall components (see
the conductivity tensor at leading order). Anyway, we shall show in the
following that the precise form we have found for the local conductivity tensor
leads to reliable physical results.

Inserting second-order contribution (\ref{transcorr2}) into continuity
Eq. (\ref{continuity}), condition (\ref{nice}) is now replaced by the
differential equation

\begin{widetext}
\begin{eqnarray}
\left( {\bm \nabla_{\bf r}} \sigma_{H} \times {\bm \nabla_{\bf r}} \Phi \right) \cdot \hat{{\bf z}}
- \frac{l_{B}^{2}}{\hbar \omega_{c}}
\sigma_{H}
\mathrm{Tr} \left\{
\left( \begin{array}{cc} 0 & -1 \\
1 & 0\end{array}\right)
\hat{{\cal H}}\left[
V_{\mathrm{eff}} \right]
\hat{{\cal H}}
\left[
\Phi \right]
\right\}
= 0,
\label{nice2}
\end{eqnarray}
\end{widetext}
where the notation $\mathrm{Tr}$ means the trace. In the neighborhood of a
critical point, which for practical convenience is taken at the origin
($x=y=0$), we have
\begin{equation}
{\bm \nabla_{\bf r}} \sigma_{H}({\bf r}) \approx
\left.
\left(
{\bf r} \cdot {\bm \nabla_{\bf r}} \right)
{\bm \nabla_{\bf r}} \sigma_{H}\right|_{{\bf 0}}.
\end{equation}
The Hessian matrices of the function $\sigma_{H}({\bf r})$ and of the function
$V_{\mathrm{eff}}({\bf r})$ being proportional at the critical point, we can
choose, without loss of generality according to the form of the Eq.
(\ref{nice2}), the $\hat{{\bf x}}$ and $\hat{{\bf y}}$ axes such that both
Hessian matrices are diagonal. This means, e.g., that $V_{\mathrm{eff}}({\bf
r})$ is expanded close to the origin as

\begin{eqnarray}
V_{\mathrm{eff}}({\bf r}) = V_{\mathrm{eff}}({\bf 0})+ a \, \frac{x^{2}}{2} +
 b \, \frac{y^{2}}{2}
\end{eqnarray}
where $a=\left. \partial_{xx}^{2}V_{\mathrm{eff}} \right|_{{\bf 0}} $ and
$b= \left. \partial_{yy}^{2} V_{\mathrm{eff}} \right|_{{\bf 0}}$.
For this situation, Eq. (\ref{nice2}) becomes then
\begin{eqnarray}
(a-b) \, \partial_{xy}^{2} \Phi
 + \frac{1}{\lambda^{2}} \left[ a x \, \partial_{y} \Phi
- b y \, \partial_{x} \Phi
\right] =0,
\label{diff}
\end{eqnarray}
with
\begin{equation}
\lambda^{2}=l_{B}^{2}
\frac{
\displaystyle \sum_{m=0}^{+ \infty}
n_{F}\left( \xi_{m}({\bf 0})\right)
}{
\displaystyle
\sum_{m=0}^{+ \infty} \hbar \omega_{c}
\left[ - n'_{F}\left( \xi_{m}({\bf 0})\right)
\right]
}.
\label{lambda}
\end{equation}
To get rid of coefficients $a$ and $b$ in the differential equation [Eq. (\ref{diff})],
it is useful to introduce the change in variables
\begin{eqnarray}
x'&=& \alpha x - \beta y \\
y'& = & \alpha x + \beta y.
\end{eqnarray}

If the critical point corresponds to a local extremum (situation with $ab>0$), we can take
\begin{eqnarray}
\alpha = \sqrt{\left|\frac{a}{a-b}\right|}, \hspace*{0.3cm}
\beta = \sqrt{\left|\frac{b}{a-b}\right|},
\end{eqnarray}
and Eq. (\ref{diff}) then reduces to
\begin{equation}
\partial_{y'y'}^{2} \Phi - \partial_{x'x'}^{2} \Phi + \frac{\epsilon}{\lambda^{2}}
\left[x' \partial_{y'} \Phi- y' \partial_{x'} \Phi \right]=0
\label{extr},
\end{equation}
where $\epsilon=+1$ if $|a|>|b|$ and $\epsilon=-1$ if $|a|<|b|$.
The general solution of Eq. (\ref{extr}) is
\begin{eqnarray}
\Phi(x',y')= \left[ A + B \, e^{-\epsilon \frac{x'y'}{\lambda^{2}}} \right]\left[C + D \left(x'^{2}+y'^{2} \right)
\right],
\end{eqnarray}
where the coefficients $A$, $B$, $C$ et $D$ are constants of integration. As
boundary conditions, we require that the electrochemical potential tends to
constant values far from the critical point. We therefore necessarily get
$D=B=0$. There is consequently no macroscopic voltage drop associated with the
crossing of a local extremum (the special case $a=b \neq 0$, which can be
readily obtained from Eq. (\ref{diff}) leads to the same result).

Now, if the critical point corresponds to a saddle-point (situation with $ab <0$), we can choose
\begin{eqnarray}
\alpha = \sqrt{\frac{a}{a-b}}, \hspace*{0.3cm}
\beta = \sqrt{\frac{b}{b-a}},
\end{eqnarray}
so that Eq. (\ref{diff}) becomes
\begin{eqnarray}
\partial_{y'y'}^{2} \Phi - \partial_{x'x'}^{2} \Phi + \frac{1}{\lambda^{2}}
\left[y' \partial_{y'} \Phi- x' \partial_{x'} \Phi \right]=0
\label{saddle}.
\end{eqnarray}
Looking for a solution with separable spatial dependences, we find that the only
solution of Eq. (\ref{saddle}) represents the product of two error step
functions in the $x'$ and $y'$ directions,
\begin{equation}
\Phi(x',y')=\left[ A + B \, \mathrm{erf}\left( \frac{x'}{ \sqrt{2} \lambda}\right) \right]
\left[ C + D \, \mathrm{erf}\left( \frac{y'}{\sqrt{2} \lambda}\right) \right]
,
\label{solsaddle}
\end{equation}
where
\begin{equation}
\mathrm{erf}(x)=\frac{2}{\sqrt{\pi}} \int_{0}^{x} e^{-t^{2}} dt .
\end{equation}
We observe with the solution (\ref{solsaddle}) that far from the saddle-point
the electrochemical potential tends to different constant values depending on
sectors. This solution is consistent with the picture of four different regions
characterized by four different electrochemical potential values with an
electrochemical drop resulting from the saddle-point crossing, which is a major
ingredient in the network models that have been developed to describe the peaks
of the longitudinal conductance in the transition regime between quantized Hall
plateaus (see, e.g., Ref. \onlinecite{Ruzin1995}). Thus, our conductivity tensor
confirms at a microscopic level the special role played by the saddle-points of
the density in the dissipative features
\cite{Fertig1987,Chalker1988,Cooper1993,Chklovskii1993,Ruzin1995,Kramer2005}.

Finally, we turn back to the condition of validity of Eq. (\ref{nice2})
which has been established under the assumption that the nonlocal term in Eq.
(\ref{transcorr}) involving the third-order derivatives of $\Phi$ play a
negligible role. Clearly, this is justified provided that $\Phi$ is smooth
enough. Considering expression (\ref{lambda}) giving the characteristic
lengthscale $\lambda$ for the spatial variations of the electrochemical
potential, we note that this assumption appears fully justified as long as the
function $n'_{F}(\xi_{m}({\bf 0}))$ remains quite small, i.e., as long as the
saddle-point filling factor is close to an integer. Conversely, we can conclude
that the nonlocal term which is associated with quantum tunneling becomes
non-negligible at low temperatures when the local chemical potential $\mu({\bf
0})=\mu^{\ast}-V_{\mathrm{eff}}({\bf 0})$ approaches a Landau level. This regime which occurs for a narrow range in
magnetic fields will be investigated in detail elsewehere.

\section{Conclusion and Perspectives
\label{Conclusion}
}

\subsection{Summary}

In summary, we have developed a systematic high magnetic field expansion,
which permits to find in a recursive way, order by order in powers of the
magnetic length $l_{B}$, Green's functions for the quantum problem of
an electron confined to a plane and subjected to a slowly-varying potential in
high magnetic fields. Using this theory, we have derived functional
quantum expressions for the local equilibrium density distribution and current
density at the first two leading orders. These expressions which contain Landau level mixing processes in a
controlled way and quantum smearing effects associated with the finite extent of
the wave function at finite magnetic fields form the starting point for future
quantitative investigations of screening effects at low temperatures in
two-dimensional disordered Hall liquids. We have checked the accuracy of our general functionals against the exact solution of a one-dimensional parabolic confining potential, demonstrating the controlled character of the theory to get equilibrium properties.
Furthermore, we have shown that our technique gives a natural and systematic access to semiclassical expansions in powers of the magnetic length of the physical observables. For example, 
we have been able to derive for the first time the semiclassical corrections of order $l_{B}^{2}$ for the local charge and current densities.

Moreover, we have proved microscopically that in high magnetic fields the
electronic system can be described within a local hydrodynamic regime and that
the electrical conduction transport takes a quasilocal form. As an important
result, we have put forward that our approximation scheme with the $l_{B}$
expansion  intrinsically captures dissipation mechanisms at the microscopic
level and accounts for quantum tunneling processes. For example, we have
derived microscopic expressions for the local conductivity tensor, which
contains both Hall and longitudinal components, the dissipative features
appearing at the order $l_{B}^{2}$, i.e., at finite magnetic fields. Furthermore,
we have established from the special form of this local conductivity tensor that
a nonzero gradient of the electrochemical potential is exclusively generated by
the saddle-points of the density distribution. A general understanding of the
transport properties at the microscopic level now seems accessible. However, the
procedure of computation of the macroscopic transport coefficients, and in
particular the treatment of nonlocal effects induced by quantum tunneling,
requires additional work, that is currently under way.

\subsection{General perspective}

Finally, we want to address a more general perspective, which is beyond the
quantum Hall effect, namely the issue of dissipation in physics and especially
in quantum mechanics. Indeed, we believe that our systematic expansion in
ascending powers of the magnetic length could shed new light on this important
issue by illustrating how the irreversible evolution of a quantum system can emerge from the consideration of
microscopic equations which are time reversal invariant. The system considered in this
paper is maybe the simplest system one can consider to answer the latter
question, since it involves only 2 degrees of freedom. Interestingly, the
magnetic field (which plays the role of a tuning parameter here) controls the
degree of mixing between these two degrees of freedom, which corresponds at
the classical level to the cyclotron motion and the guiding-center motion. In
high magnetic fields and for a smooth arbitrary potential, this mixing becomes
very weak as a result of the strongly different timescales associated with the
two kinds of motion. The system even becomes  dynamically
integrable in the strict limit of infinite magnetic fields when the mixing between the two degrees of freedom is
no more possible. Therefore, the high but finite magnetic field regime can be
associated with a classical regime of soft chaos. At the quantum-mechanical
level, it is clear that the quantization of the kinetic orbital motion which introduces
robustness (in the sense that it  considerably constrains the possible variations
of the orbital motion) renders this exchange between the 2 degrees of freedom
even much more ineffective. We thus expect that the quantum system is somehow even
closer to integrability than the classical one.

In classical chaotic systems (this is, for example, the case for the
present disordered system in low magnetic fields), irreversibility and dissipation are
often associated with the technical impossibility to fully describe the trajectories as a result of complicated mixing mechanisms between
the degrees of freedom. This complexity is then transposed in
terms of a stochastic description, thus expressing a loss of information. We
have shown that in high magnetic fields it is not required to average over the
disorder configuration in order to find an analytical approximate  solution to the quantum problem,
contrary to the situation at low magnetic fields. Moreover, we have noticed that time irreversibility has nevertheless been introduced at some stage of the derivation,
%
 since our high magnetic field theory accounts for dissipation features related to time-decaying states. It turns out
from first considerations that the dissipation involves Landau level mixing processes and arises from the conjunction of  local quantum fluctuations with a local dynamical instability taking place at the saddle-points of the local equilibrium density (a saddle-point is necessarily characterized by stable and unstable directions which can be defined in an obvious manner). 
Interestingly,  in low magnetic fields  the electrical conduction is also directly related to another  instability mechanism which is realized by the sensitivity to the initial condition characterizing the chaotic systems. 
In brief, insights in this general
perspective of understanding the emergence of dissipation in quantum-mechanical
systems could be gained from closer investigations of the high magnetic field expansion developed in the present work.

\section*{Acknowledgement}

We thank V.P. Mineev for drawing our attention to Ref. \onlinecite{Geller1994}. Stimulating discussions with R. Citro, M. Houzet, V. Rossetto, S. Skipetrov, and D. Venturelli are also gratefully acknowledged.

\appendix
\begin{widetext}
\section{Off-diagonal elements of third-order Green's functions \label{apg3}}

In this Appendix, we provide the detail for the derivation of the elements
$\delta_{m_{1},m_{2} \pm 1}$ of third-order Green's function $g^{(3)}$, that
are needed for the computation of the second-order current density performed in
Sec. \ref{section4} and Appendix \ref{appj}. The function $g^{(3)}$ obeys the equation

\begin{eqnarray}
\omega_{m_{1}} g^{(3)}_{m_{1};m_{2}}({\bf R})
& = & v^{(3)}_{m_{1};m_{2}} ({\bf R})g^{(0)}_{m_{2};m_{2}}({\bf R})+ \sum_{m_{3}}
\left[
v^{(2)}_{m_{1};m_{3}} ({\bf R}) g^{(1)}_{m_{3};m_{2}}({\bf R})
+
v^{(1)}_{m_{1};m_{3}} ({\bf R}) g^{(2)}_{m_{3};m_{2}}({\bf R})
\right] \nonumber \\
&&
+(\partial_{X}-i\partial_{Y})V({\bf R}) (\partial_{X}+i\partial_{Y}) g^{(1)}_{m_{1};m_{2}}({\bf R})
+
(\partial_{X}-i\partial_{Y})v^{(1)}_{m_{1};m_{2}}({\bf R}) (\partial_{X}+i\partial_{Y}) g^{(0)}_{m_{2};m_{2}}({\bf R})
.
\end{eqnarray}
Here is a list of these numerous components $\delta_{m_{1},m_{2} \pm 1}$ of $g^{(3)}_{m_{1};m_{2}} ({\bf R})$ arising
\begin{itemize}
\item from the combination $v^{(3)} g^{(0)}$:
\begin{eqnarray}
\frac{1}{2 \omega_{m_{1}}\omega_{m_{2}}} \left[ (m_{1}+1)
\sqrt{m_{1}} \, \delta_{m_{1},m_{2}+1} (\partial_{X}-i \partial_{Y})+(m_{2}+1) \sqrt{m_{2}}
\delta_{m_{1}+1,m_{2}} (\partial_{X}+i \partial_{Y})\right]\Delta V ,
\label{A2}
\end{eqnarray}

\item from the combination $v^{(2)} g^{(1)}$:
\begin{eqnarray}
\frac{m_{1}+1}{\omega^{2}_{m_{1}} \omega_{m_{2}}} \Delta_{\bf R} V
\left[ \sqrt{m_{1}} \, \delta_{m_{1},m_{2}+1} (\partial_{X}-i \partial_{Y})+
 \sqrt{m_{2}} \, \delta_{m_{1}+1,m_{2}} (\partial_{X}+i \partial_{Y})
\right]V
\nonumber \\
+\frac{m_{2} \sqrt{m_{1}}}{2\omega_{m_{2}-1} \omega_{m_{2}}\omega_{m_{2}+1} } \, \delta_{m_{1},m_{2}+1}
\left[(\partial_{X}-i\partial_{Y})^{2}V
\right]
(\partial_{X}+i\partial_{Y})V
\nonumber \\
+\frac{(m_{2}+1) \sqrt{m_{2}}}{2 \omega_{m_{2}-1} \omega_{m_{2}}\omega_{m_{2}+1}} \, \delta_{m_{1}+1,m_{2}}
\left[(\partial_{X}+i\partial_{Y})^{2}V
\right]
(\partial_{X}-i\partial_{Y})V
,
\label{A3}
\end{eqnarray}

\item from the combination $v^{(1)} g^{(2)}$:

\begin{eqnarray}
\frac{m_{2}+1}{\omega_{m_{1}} \omega_{m_{2}}^{2} } \Delta_{\bf R} V
\left[
\sqrt{m_{1}} \, \delta_{m_{1},m_{2}+1} (\partial_{X}-i\partial_{Y})+
\sqrt{m_{2}} \, \delta_{m_{1}+1,m_{2}} (\partial_{X}+i\partial_{Y})
\right]V
\nonumber
\\
+ \left[
\frac{m_{2}+1}{\omega_{m_{2}+1}}
+ \frac{m_{2}}{\omega_{m_{2}-1}}
+\frac{1}{\omega_{m_{2}}}
\right]
\frac{ \left|{\bm \nabla_{\bf R}}V \right|^{2} }{\omega_{m_{2}}^{2} \omega_{m_{1}}}
\left[
\sqrt{m_{1}} \, \delta_{m_{1},m_{2}+1} (\partial_{X}-i\partial_{Y})+
\sqrt{m_{2}} \, \delta_{m_{1}+1,m_{2}} (\partial_{X}+i\partial_{Y})
\right]V
 \nonumber \\
+\frac{m_{1} \sqrt{m_{2}} }{2 \omega_{m_{1}-1}\omega_{m_{1}} \omega_{m_{1}+1}}
\,
\delta_{m_{1}+1,m_{2}}
\left[
(\partial_{X}-i \partial_{Y})V
\right]
\left[
(\partial_{X}+i \partial_{Y})^{2}
V
\right] \nonumber
\\
+\frac{(m_{1}+1) \sqrt{m_{1}} }{2 \omega_{m_{1}-1}\omega_{m_{1}} \omega_{m_{1}+1}}
\,
\delta_{m_{1},m_{2}+1}
\left[
(\partial_{X}+i \partial_{Y})V
\right]
\left[
(\partial_{X}-i \partial_{Y})^{2}
V
\right]
 \nonumber
\\
+
\frac{ m_{1} \sqrt{m_{2}}}{\omega_{m_{1}-1}\omega_{m_{1}}^{2} \omega_{m_{1}+1} } \,
\delta_{m_{1}+1,m_{2}}
\left[
(\partial_{X}-i \partial_{Y})V
\right]
\left[
(\partial_{X}+i\partial_{Y})V
\right]^{2}
\nonumber
\\
+
\frac{(m_{1}+1) \sqrt{m_{1}}}{\omega_{m_{1}-1}\omega_{m_{1}}^{2} \omega_{m_{1}+1} } \,
\delta_{m_{1},m_{2}+1}
\left[
(\partial_{X}+i \partial_{Y})V
\right]
\left[
(\partial_{X}-i\partial_{Y})V
\right]^{2}
,
\label{A4}
\end{eqnarray}

\item and from the combinations $(\partial_{X}-i\partial_{Y})v (\partial_{X}+i \partial_{Y}) g$:

\begin{eqnarray}
\left|{\bm \nabla_{\bf R}}V \right|^{2}
\left[ \frac{1}{\omega_{m_{1}}^{3} \omega_{m_{2}}}
+\frac{1}{\omega_{m_{1}}^{2} \omega_{m_{2}}^{2}}
\right]
\left[
\sqrt{m_{1}}(\partial_{X}-i\partial_{Y}) \delta_{m_{1},m_{2}+1}
+
\sqrt{m_{2}}(\partial_{X}+i\partial_{Y}) \delta_{m_{1}+1,m_{2}}
\right]V
\nonumber \\
+ \frac{\Delta_{\bf R} V }{\omega_{m_{1}}\omega_{m_{2}}}
\left[
\frac{\sqrt{m_{1}}}{\omega_{m_{1}}}
(\partial_{X}-i\partial_{Y}) \delta_{m_{1},m_{2}+1}
+
\frac{\sqrt{m_{2}}}{\omega_{m_{2}}}
(\partial_{X}+i\partial_{Y}) \delta_{m_{1}+1,m_{2}}
\right]V
\nonumber \\
+
\frac{\sqrt{m_{1}}}{\omega_{m_{1}} \omega_{m_{2}}^{2} } \, \delta_{m_{1},m_{2}+1}
\left[(\partial_{X}-i\partial_{Y})^{2}V\right] \left[(\partial_{X}+i \partial_{Y})V \right]
\nonumber \\
+
\frac{ \sqrt{m_{2}} }{ \omega_{m_{1}}^{2} \omega_{m_{2}}} \, \delta_{m_{1}+1,m_{2}}
\left[(\partial_{X}+i\partial_{Y})^{2}V\right] \left[(\partial_{X}-i \partial_{Y})V \right]
.
\label{A5}
\end{eqnarray}

\end{itemize}
Regrouping the terms of the same form, different contributions (\ref{A2})-(\ref{A5})
to the components $\delta_{m_{1},m_{2} \pm 1}$ of $g^{(3)}$ are rearranged as

\begin{itemize}
\item terms with $\Delta_{\bf R} (\partial_{X} \pm i \partial_{Y})V$:

\begin{eqnarray}
\frac{1}{2} \Delta_{\bf R} \left[
(m_{1}+1) \sqrt{m_{1}} \, \delta_{m_{1},m_{2}+1} (\partial_{X}-i\partial_{Y})
+
(m_{2}+1) \sqrt{m_{2}} \, \delta_{m_{1}+1,m_{2}} (\partial_{X}+i\partial_{Y})
\right]
V/\omega_{m_{1}} \omega_{m_{2}},
\label{g3a}
\end{eqnarray}

\item terms with $\Delta_{\bf R} V (\partial_{X}\pm i\partial_{Y})V$:

\begin{eqnarray}
\Delta_{\bf R} V
\left[
\frac{m_{2}+1}{\omega_{m_{1}}\omega_{m_{2}}^{2}}
+
\frac{m_{1}+1}{\omega_{m_{2}}\omega_{m_{1}}^{2}}
\right]
\left[
\sqrt{m_{1}} \, \delta_{m_{1},m_{2}+1} (\partial_{X}-i\partial_{Y})
+
\sqrt{m_{2}} \, \delta_{m_{1}+1,m_{2}} (\partial_{X}+i\partial_{Y})
\right]
V
\nonumber \\
+ \frac{\Delta_{\bf R} V }{\omega_{m_{1}}\omega_{m_{2}}}
\left[
\frac{\sqrt{m_{1}}}{\omega_{m_{1}}}
(\partial_{X}-i\partial_{Y}) \delta_{m_{1},m_{2}+1}
+
\frac{\sqrt{m_{2}}}{\omega_{m_{2}}}
(\partial_{X}+i\partial_{Y}) \delta_{m_{1}+1,m_{2}}
\right]V
,
\end{eqnarray}

\item terms with $[(\partial_{X} \pm i \partial_{Y})^{2} V]
(\partial_{X}\mp i\partial_{Y})V$:

\begin{eqnarray}
\frac{1}{2}
\left\{
\left[
\frac{m_{2}}{\omega_{m_{2}-1}\omega_{m_{2}}\omega_{m_{2}+1}}
+
\frac{m_{1}+1}{\omega_{m_{1}-1}\omega_{m_{1}}\omega_{m_{1}+1}}
+\frac{2}{\omega_{m_{1}} \omega_{m_{2}}^{2}}
\right]
\sqrt{m_{1}} \, \delta_{m_{1},m_{2}+1}
\left[
 (\partial_{X}-i\partial_{Y})^{2}V
\right]
\left[
(\partial_{X}+i \partial_{Y})V
\right]
\nonumber \right.
\\
+
\left.
\left[
\frac{m_{1}}{\omega_{m_{1}-1}\omega_{m_{1}}\omega_{m_{1}+1}}
+
\frac{m_{2}+1}{\omega_{m_{2}-1}\omega_{m_{2}}\omega_{m_{2}+1}}
+\frac{2}{\omega_{m_{1}}^{2} \omega_{m_{2}}}
\right]
\sqrt{m_{2}} \, \delta_{m_{1}+1,m_{2}}
\left[
 (\partial_{X}+i\partial_{Y})^{2}V
\right]
\left[
(\partial_{X}-i \partial_{Y})V
\right]
\right\}
,
\end{eqnarray}

\item terms with $\left|{\bm \nabla_{\bf R}} V \right|^{2}
(\partial_{X}\pm i\partial_{Y})V$:
\begin{eqnarray}
\frac{\left|
{\bm \nabla_{\bf R}}V \right|^{2}}
{\omega_{m_{1}}\omega_{m_{2}}}
\left\{
\left[
\frac{m_{1}+1}{\omega_{m_{1}} \omega_{m_{1}+1}}
+
\frac{m_{2}}{\omega_{m_{2}-1}\omega_{m_{2}}}
+ \frac{m_{1}+1}{\omega_{m_{1}}\omega_{m_{2}}}
+ \frac{1}{\omega_{m_{1}}^{2}}
+ \frac{1}{\omega_{m_{2}}^{2}}
\right]
\sqrt{m_{1}} \, \delta_{m_{1},m_{2}+1}
\left[
(\partial_{X}-i \partial_{Y})V
\right]
\nonumber \right.
\\
+
\left.
\left[
\frac{m_{2}+1}{\omega_{m_{2}} \omega_{m_{2}+1}}
+
\frac{m_{1}}{\omega_{m_{1}-1}\omega_{m_{1}}}
+ \frac{m_{2}+1}{\omega_{m_{1}} \omega_{m_{2}}}
+ \frac{1}{\omega_{m_{1}}^{2}}
+ \frac{1}{\omega_{m_{2}}^{2}}
\right]
\sqrt{m_{2}} \, \delta_{m_{1}+1,m_{2}}
\left[
(\partial_{X}+i \partial_{Y})V
\right]
\right\}
.
\label{g3d}
\end{eqnarray}

\end{itemize}

\section{Proof of useful relations \label{approofs}}

In this appendix we prove identities (\ref{practical}), and (\ref{70})-(\ref{71}).
First, with the help of Eq. (\ref{60}) we can find that
\begin{eqnarray}
\left(
\partial_{x}- i \partial_{y}
\right)
\left|
 \Psi_{p,{\bf R}}({\bf r})
\right|^{2}
=\frac{\sqrt{2}}{l_{B}}
\left(
\sqrt{p} \, \Psi_{p,{\bf R}}^{\ast}({\bf r}) \Psi_{p-1,{\bf R}}({\bf r})
- \sqrt{p+1} \,
 \Psi_{p+1,{\bf R}}^{\ast}({\bf r})
 \Psi_{p,{\bf R}}({\bf r})
\right),
\label{begin}
\end{eqnarray}
what defines in a recursive way the combination $\sqrt{p+1} \, \Psi_{p+1,{\bf R}}^{\ast}({\bf r})
 \Psi_{p,{\bf R}}({\bf r})$. From this relation (\ref{begin}), it is
then straightforward to obtain identity (\ref{practical}).

Using Eq. (\ref{60}), it can be readily established that
\begin{eqnarray}
\left(\partial_{x}-i\partial_{y} \right)
\left\{
 \Psi_{m,{\bf R}}({\bf r}) \Psi_{m+1,{\bf R}}^{\ast}({\bf r})
\right\}
=\frac{\sqrt{2}}{l_{B}} \left[
\sqrt{m} \, \Psi_{m-1,{\bf R}}({\bf r}) \Psi_{m+1,{\bf R}}^{\ast}({\bf r})- \sqrt{m+2} \,
\Psi_{m,{\bf R}}({\bf r}) \Psi_{m+2,{\bf R}}^{\ast}({\bf r})
\right].
\label{Bbis}
\end{eqnarray}
From this relation (\ref{Bbis}), we deduce that
\begin{eqnarray}
\sqrt{m+2} \,
\Psi_{m+2,{\bf R}}^{\ast}({\bf r})\Psi_{m,{\bf R}}({\bf r})
=
\sqrt{m} \, \Psi_{m+1,{\bf R}}^{\ast} ({\bf r}) \Psi_{m-1,{\bf R}}({\bf r})
- \frac{l_{B}}{\sqrt{2}}
\left( \partial_{x}-i \partial_{y}\right)
\left\{
 \Psi_{m+1,{\bf R}}^{\ast}({\bf r }) \Psi_{m,{\bf R}}({\bf r})
\right\}
.
\label{B2}
\end{eqnarray}
Multiplying Eq. (\ref{B2}) by $\sqrt{m+1}$, we get a recursive relation which yields
\begin{eqnarray}
\sqrt{m+1}\sqrt{m+2} \,
\Psi_{m+2,{\bf R}}^{\ast}({\bf r}) \Psi_{m,{\bf R}}({\bf r})
=
- \frac{l_{B}}{\sqrt{2}}
\left( \partial_{x}-i \partial_{y}\right)
\sum_{p=0}^{m}
\sqrt{p+1} \, \Psi_{p+1,{\bf R}}^{\ast}({\bf r}) \Psi_{p,{\bf R}}({\bf r})
.
\end{eqnarray}
Finally, using identity (\ref{practical}), we get the result Eq. (\ref{70}).

From Eq. (\ref{60}), we can get
\begin{eqnarray}
\Delta_{{\bf r}}
\left| \Psi_{p,{\bf R}}({\bf r})\right|^{2}
=
\frac{2 }{l_{B}^{2}}
\left\{
(p+1)\left| \Psi_{p+1,{\bf R}}({\bf r})\right|^{2}
+
p \left| \Psi_{p-1, {\bf R}}({\bf r})\right|^{2}
-(2p+1)\left| \Psi_{p,{\bf R}}({\bf r})\right|^{2}
\right\}.
\end{eqnarray}
Therefore, we can write
\begin{eqnarray}
\sum_{p=0}^{m} (m+1-p)
\Delta_{{\bf r}}
\left| \Psi_{p,{\bf R}}({\bf r})\right|^{2}
 & =&
\frac{2 }{l_{B}^{2}}
\sum_{p=0}^{m} (m+1-p)
\left\{
(p+1)\left| \Psi_{p+1,{\bf R}}({\bf r})\right|^{2}
+
p \left| \Psi_{p-1,{\bf R}}({\bf r})\right|^{2}
-(2p+1)\left| \Psi_{p, {\bf R}}({\bf r})\right|^{2}
\right\}
\nonumber
\\
&=&
\frac{2 }{l_{B}^{2}}
\left(
(m+1) \left| \Psi_{m+1,{\bf R}}({\bf r})\right|^{2}
-\sum_{p=0}^{m} \left| \Psi_{p,{\bf R}}({\bf r})\right|^{2}
\right)
,
\end{eqnarray}
which proves identity (\ref{71}).

\section{Calculation of the electronic current at order $l_B^2$}
\label{appj}

In this appendix we present the detailed derivation of the quantum (Appendix \ref{quantj}) and semiclassical (Appendix \ref{semj}) expressions for the electronic current density up to order $l_{B}^{2}$.

\subsection{Quantum expressions for the current
\label{quantj}
}

\subsubsection{Contribution from $g^{(1)}$}
First-order vortex Green's function has the total contribution to the current
given by formula~(\ref{j1}), from which the leading-order term was extracted
in Eq.~(\ref{j1fin}). We express here the formula (\ref{j1}) in a form that makes explicit its leading and subdominant contributions.
Using the identities proven in Appendix \ref{approofs}
\begin{eqnarray}
\sqrt{m+1}\sqrt{m+2} \,
 \Psi_{m+2,{\bf R}}^{\ast}({\bf r}) \Psi_{m,{\bf R}}({\bf r})
&= &
\left[-
\frac{l_{B}}{\sqrt{2}}\left(\partial_{x}-i\partial_{y} \right)
\right]^{2}
\sum_{p=0}^{m}(m+1-p) \left|\Psi_{p,{\bf R}}({\bf r}) \right|^{2},
\label{70}
\\
\frac{l_{B}^{2}}{2}\Delta_{{\bf r}}
\sum_{p=0}^{m}(m+1-p) \left|\Psi_{p,{\bf R}}({\bf r}) \right|^{2}
&=&
(m+1) \left| \Psi_{m+1,{\bf R}}({\bf r})\right|^{2}
-
\sum_{p=0}^{m}
 \left| \Psi_{p,{\bf R}}({\bf r})\right|^{2}
,
\label{71}
\end{eqnarray}
we can rewrite the combination $\Psi_{m,{\bf R}}({\bf r}) \Psi_{m+2,{\bf
R}}^{\ast}({\bf r}) $ of vortex wave functions appearing in Eq. (\ref{j1}) in the
following way:
\begin{eqnarray}
\sqrt{m+1}\sqrt{m+2}\left(
\begin{array}{c}
\mathrm{Im} \\
\mathrm{Re}
\end{array}
\right)
(\partial_{X} V+i \partial_{Y}V)
\Psi_{m,{\bf R}}({\bf r})
\Psi_{m+2,{\bf R}}^{\ast}({\bf r})
=
l_{B}^{2}
\hat{{\bf z}}
\times \left( {\bm \nabla}_{{\bf R}} V \cdot {\bm \nabla}_{{\bf r}} \right)
{\bm \nabla}_{{\bf r}}
\sum_{p=0}^{m}(m+1-p) \left| \Psi_{p,{\bf R}}({\bf r})\right|^{2}
\nonumber \\
+
\hat{{\bf z}}
\times {\bm \nabla}_{{\bf R}}V \left(
\sum_{p=0}^{m}
 \left| \Psi_{p,{\bf R}}({\bf r})\right|^{2}
-(m+1) \left| \Psi_{m+1,{\bf R}}({\bf r})\right|^{2}
\right).
\label{mieux}
\end{eqnarray}
Inserting expressions (\ref{utile}) and (\ref{mieux}) in Eq. (\ref{j1}), we then express the current density as
\begin{eqnarray}
{\bf j}^{(1)}({\bf r}) &=& 
\frac{e \hbar }{m^{\ast}}
\int\!\!\! \frac{d^{2}{\bf R}}{2 \pi l_{B}^{2}}
\sum_{m=0}^{+\infty}
\frac{
n_{F}(\xi_{m+1}({\bf R})) -n_{F}(\xi_{m}({\bf R}))}{\hbar \omega_{c}}
\left(
\sum_{p=0}^{m}
\left| \Psi_{p,{\bf R}}({\bf r})\right|^{2}
 \hat{{\bf z}} \times {\bm \nabla}_{{\bf R}} V({\bf R})
\right.
\nonumber \\
&&
+ l_{B}^{2}
\left.
 \hat{{\bf z}} \times \left[ {\bm \nabla}_{{\bf R}} V({\bf R}) \cdot {\bm \nabla}_{{\bf r}}\right]
 {\bm \nabla}_{{\bf r}}
 \sum_{p=0}^{m}(m+1/2-p)
\left| \Psi_{p,{\bf R}}({\bf r})\right|^{2}
 \right),
\label{presque}
\end{eqnarray}
where we have used
\begin{equation}
{\bm \nabla}_{{\bf r}}\left( {\bm \nabla}_{{\bf R}}
V \cdot {\bm \nabla}_{{\bf r}}\left| \Psi_{p,{\bf R}}({\bf r})\right|^{2}\right)
=
\left[ {\bm \nabla}_{{\bf R}} V \cdot {\bm \nabla}_{{\bf r}}\right]
 {\bm \nabla}_{{\bf r}}
\left(
\left| \Psi_{p,{\bf R}}({\bf r})\right|^{2}\right)
.
\end{equation}
After a straightforward simplification of expression (\ref{presque}), the current density ${\bf j}^{(1)}$ can finally be divided into a leading contribution given by Eq. (\ref{j1fin}) and a subdominant contribution which reads
\begin{equation}
{\bf j}^{(1)}_\mathrm{sub}({\bf r}) =
\frac{e}{h}
\int\!\!\! d^{2}{\bf R}
\sum_{m=0}^{+\infty}
n_{F}(\xi_{m}({\bf R})) \, 
l_{B}^{2}
 \left[ {\bm \nabla}_{{\bf R}} V({\bf R}) \cdot {\bm \nabla}_{{\bf r}}\right]
 {\bm \nabla}_{{\bf r}}
\left(
 \sum_{p=0}^{m}
\left| \Psi_{p,{\bf R}}({\bf r})\right|^{2}
-
\frac{\left| \Psi_{m,{\bf R}}({\bf r})\right|^{2}}{2}
\right)
\times \hat{{\bf z}}
.
\label{j1sub}
\end{equation}

\subsubsection{Contribution from $g^{(2)}$}
Second-order vortex Green's function~(\ref{g2}) contains diagonal elements
($m=m'$), which contribute within the first and second terms in the rhs
of Eq. (\ref{simp}). It has also off-diagonal elements $\delta_{m,m' \pm 2}$,
which combine with the second term of the rhs of Eq. (\ref{simp}),
to give terms involving wave functions with adjacent Landau levels (wave functions with
a Landau index difference of 3 are also obtained, but these contribute to the
current at order $l_B^4$ and will be discarded).
After inspection, the contribution from the function $g^{(0)}$ appearing with the
term $k=1$ in Eq. (\ref{simp}) combine very naturally with these terms from
$g^{(2)}$, so that the starting expression reads:
\begin{eqnarray}
{\bf j}^{(2)}({\bf r}) =
\frac{e \hbar }{ 2 m^{\ast}} \hat{{\bf z}} \times {\bm \nabla}_{{\bf r}} \rho^{(2)}({\bf r})
+
\frac{e \hbar }{m^{\ast}}
\int \!\!\! \frac{d\omega}{2 \pi}
\int\!\!\! \frac{d^{2}{\bf R}}{2 \pi l_{B}^{2}}
\sum_{m=0}^{+\infty}
\frac{l_{B}}{\sqrt{2}}
 \sqrt{m+1}
\,
\left(\begin{array}{c} \mathrm{Re} \\ \mathrm{Im}\end{array} \right)
\left\{
\Psi_{m+2,{\bf R}}^{\ast}({\bf r}) \Psi_{m+1,{\bf R}}({\bf r}) g^{(2)<}_{m;m+2}({\bf R},\omega)
\right.
\nonumber
\\
\left.
+
\Psi_{m+1,{\bf R}}({\bf r}) \Psi_{m,{\bf R}}^{\ast}({\bf r})
\left[
g^{(2)<}_{m;m}({\bf R},\omega)
-
\Delta_{{\bf R}} g^{(0)<}_{m;m}({\bf R},\omega)
\right]
\right\}
.
\label{j2prem}
\end{eqnarray}
After using Eqs. (\ref{practical}) and (\ref{rho2diag}), and performing the
remaining energy integration, we can rewrite expression (\ref{j2prem}) as
\begin{eqnarray}
{\bf j}^{(2)}({\bf r}) &= &
 \frac{e}{h} \,
\int\!\!\! d^{2}{\bf R}
\sum_{m=0}^{+\infty}
\frac{l_{B}^{2}}{2}
\left\{
\left(
n'_{F}(\xi_{m}({\bf R}))\left[ m \Delta_{\bf R} V - \frac{\left| {\bm \nabla_{\bf R}} V\right|^{2}}{\hbar \omega_{c}}\right]
- n''_{F}(\xi_{m}({\bf R})) \frac{\left| {\bm \nabla_{\bf R}} V\right|^{2}}{2}
+ \frac{\left|{\bm \nabla_{\bf R}} V \right|^{2}}{(\hbar \omega_{c})^{2}}
\left[m \, n_{F}\left( \xi_{m-1}({\bf R})\right)
\right.
\right.
\nonumber
\right.
\nonumber \\
&&
\left.
 \left.
+
(m+1) n_{F}\left( \xi_{m+1}({\bf R})\right)
-(2m+1) n_{F}\left( \xi_{m}({\bf R})\right)
\right]
\frac{}{}
\right)
\hbar \omega_{c}
{\bm \nabla_{{\bf r}}} \left[ \sum_{p=0}^{m}
 \left| \Psi_{p,{\bf R}}({\bf r})\right|^{2}
-\frac{ \left| \Psi_{m,{\bf R}}({\bf r})\right|^{2}}{2}
\right]
 \nonumber
\\
&&
+
(m+1)
\left\{
\left[n_{F}(\xi_{m+2}({\bf R})) + n_{F}(\xi_{m}({\bf R})) - 2n_{F}(\xi_{m+1}({\bf R}))\right]
\left[
\frac{ {\bm \nabla_{\bf R}} V \left( {\bm \nabla_{\bf R}} V
\cdot {\bm \nabla}_{{\bf r}} \right)
}{\hbar \omega_{c}}
-\frac{\left|{\bm \nabla_{\bf R}}V \right|^{2}}{2 \hbar \omega_{c}} {\bm \nabla}_{{\bf r}}
\right]
\nonumber
\right.
\\
&&
\left. \left.
+
\left[
n_{F}(\xi_{m+2}({\bf R}))- n_{F}(\xi_{m}({\bf R}))
\right]
\left[
\frac{
\left( {\bm \nabla}_{{\bf r}} \cdot {\bm \nabla_{\bf R}} \right) {\bm \nabla_{\bf R}} V
}{2}
- \frac{\Delta V}{4} {\bm \nabla}_{{\bf r}}
\right]
\right\}
\sum_{p=0}^{m+1} \left| \Psi_{p,{\bf R}}({\bf r})\right|^{2}
\right\}
\times
 \hat{{\bf z}}
.
\label{j2nonlocal}
\end{eqnarray}

\subsubsection{Contribution from $g^{(3)}$}
Finally, there exist second-order contributions to the current density coming
from the elements of third-order Green's function $g^{(3)}$ which couple
adjacent Landau levels [this contribution arises from the second term in the
rhs of Eq. (\ref{simp})]. Similar to the previous calculation, these
recombine nicely with the contribution from the function $g^{(1)}$ associated
with the term $k=1$ in Eq. (\ref{simp}). Our starting expression thus reads
\begin{eqnarray}
{\bf j}^{(3)}({\bf r}) =
\frac{e \hbar }{m^{\ast}}
\int \!\!\! \frac{d\omega}{2 \pi}
\int\!\!\! \frac{d^{2}{\bf R}}{2 \pi l_{B}^{2}}
\sum_{m=0}^{+\infty}
\frac{l_{B}^{2}}{2}
\sqrt{m} \left| \Psi_{m,{\bf R}}({\bf r}) \right|^{2}
\left(\begin{array}{c} \mathrm{Im} \\ \mathrm{Re}\end{array} \right)
\left[
i
\Delta_{{\bf R}} g^{(1)<}_{m;m-1}({\bf R},\omega)
-ig^{(3)<}_{m;m-1}({\bf R},\omega)
\right]
.
\end{eqnarray}
Inserting the explicit expressions for the first- and third-order Green's functions
[expressions (\ref{g1}) and (\ref{g3a})-(\ref{g3d})] and performing the
integration over the energy $\omega$, we finally find after tedious
calculations
\begin{eqnarray}
{\bf j}^{(3)}({\bf r}) &=&
\frac{e}{h}
\hat{{\bf z}} \times
\int\!\!\! d^{2}{\bf R}
\sum_{m=0}^{+\infty}
\frac{l_{B}^{2}}{2} m \left| \Psi_{m,{\bf R}}({\bf r}) \right|^{2}
\left\{
\frac{m+1}{2}\left[
 n_{F}(\xi_{m}({\bf R})) -n_{F}(\xi_{m-1}({\bf R}))
\right] \Delta_{\bf R} {\bm \nabla_{\bf R}}V
\nonumber
\right.
\\
&&
+ \left[
(m+2)n'_{F}(\xi_{m}({\bf R}))
-m \, n'_{F}(\xi_{m-1}({\bf R}))
+\frac{2}{\hbar \omega_{c}}
\left[
 n_{F}(\xi_{m-1}({\bf R})) -n_{F}(\xi_{m}({\bf R}))
\right]
\right] \Delta_{\bf R} V {\bm \nabla_{\bf R}}V
\nonumber \\
&&
+
\Delta \left\{
\left[
 n_{F}(\xi_{m-1}({\bf R})) -n_{F}(\xi_{m}({\bf R}))
\right]{\bm \nabla_{\bf R}}V \right\}
+\frac{1}{4 \hbar \omega_{c}}
 \left[ \Delta_{\bf R} V {\bm \nabla_{\bf R}}V - {\bm \nabla_{\bf R}}
\left\{
\left|{\bm \nabla_{\bf R}} V \right|^{2}
\right\}\right]
\nonumber
\\
&&
\times
\left[
4 \hbar \omega_{c} n'_{F}(\xi_{m-1}({\bf R}))
+
(m+1) \left[ n_{F}(\xi_{m-1}({\bf R}))
-
 n_{F}(\xi_{m+1}({\bf R})) \right]
+(m-1)
\left[
n_{F}(\xi_{m}({\bf R}))
- n_{F}(\xi_{m-2}({\bf R}))
\right]
\right]
\nonumber
\\
&&
+
\left[ \frac{}{}
(m+1) n_{F}(\xi_{m+1}({\bf R}))-
(m-1) n_{F}(\xi_{m-2}({\bf R}))
+
(3m-1) n_{F}(\xi_{m-1}({\bf R}))
-
(3m+1) n_{F}(\xi_{m}({\bf R}))
\right.
\nonumber
\\
&&
\left. \left.
+
\frac{}{}
(\hbar \omega_{c}) ^{2}
\left[
 n''_{F}(\xi_{m}({\bf R}))
-
 n''_{F}(\xi_{m-1}({\bf R}))
\right]
+2 \hbar \omega_{c}
\left[
 n'_{F}(\xi_{m-1}({\bf R}))
-
 n'_{F}(\xi_{m}({\bf R}))
\right]
\right]
\frac{ \left|{\bm \nabla_{\bf R} }V \right|^{2} {\bm \nabla_{\bf R} V}}{2 (\hbar \omega_{c})^{2}}
\right\}
.
\label{j3nonlocal}
\end{eqnarray}

\subsection{Semiclassical expressions for the current
\label{semj}
}

The second-order contributions involve several terms according to their
different possible origins. A first term comes with the expansion of the
density-gradient contribution ${\bf j}^{(0)}$ [Eq. (\ref{j0gen})]
\begin{eqnarray}
{\bf j}^{(0)}({\bf r})
=
\frac{e}{h}
 \frac{l_{B}^{2}}{2}
\sum_{m=0}^{+ \infty}
\frac{(m+1)^{2}}{2} \hbar \omega_{c} \, \Delta_{\bf r} {\bm \nabla_{\bf r}}
\left[ n_{F}(\xi_{m}({\bf r})) \right] \times
\hat{{\bf z}},
\label{j0local}
\end{eqnarray}
where we have used Eq. (\ref{general}). After making two integration by parts
and using Eq. (\ref{general}), the second-order term for the current density
arising from Green's function $g^{(1)}$ [Eqs. (\ref{j1fin}) and (\ref{j1sub})] takes
the form
\begin{eqnarray}
{\bf j}^{(1)}({\bf r})
=
\frac{e}{h}
\frac{l_{B}^{2}}{2}
\sum_{m=0}^{+ \infty}
(3m+2) \Delta_{\bf r} \left[ n_{F}(\xi_{m}({\bf r})) {\bm \nabla_{\bf r}} V \right]
 \times
\hat{{\bf z}}
.
\label{j1local}
\end{eqnarray}
Second-order terms brought by the contribution ${\bf j}^{(2)}$ are written as

\begin{eqnarray}
{\bf j}^{(2)}({\bf r})
=
\frac{e}{h}
\frac{l_{B}^{2}}{2}
\sum_{m=0}^{+ \infty}
\left\{
\left( m+\frac{1}{2}\right) \hbar \omega_{c}
{\bm \nabla_{\bf r}}
\left[
n'_{F}(\xi_{m}({\bf r}))\left( m \Delta_{\bf r} V - \frac{\left| {\bm \nabla_{\bf r}} V \right|^{2}}{\hbar \omega_{c}}
\right)
-
n''_{F}(\xi_{m}({\bf r}))
\frac{\left| {\bm \nabla_{\bf r}} V \right|^{2}}{2}
\right] \times \hat{{\bf z}}
\nonumber
\right.
\\
\left.
+
\frac{2}{\hbar \omega_{c}}
\left[
n_{F}(\xi_{m}({\bf r}))
\left(
{\bm \nabla_{\bf r} }V \cdot {\bm \nabla_{\bf r}}
\right)
{\bm \nabla_{\bf r}} V
+
\left(
{\bm \nabla_{\bf r}} \cdot \left\{
n_{F}(\xi_{m}({\bf r}))
{\bm \nabla_{\bf r}}V
\right\} \right)
{\bm \nabla_{\bf r}} V
\right]
\times \hat{{\bf z}}
\nonumber
\right.
\\
\left.
+
\left( m+\frac{1}{2}\right)
\left[
\Delta_{\bf r} V {\bm \nabla_{\bf r}} n_{F}(\xi_{m}({\bf r}))
- n_{F}(\xi_{m}({\bf r})) \Delta_{\bf r} {\bm \nabla_{\bf r}} V
-2
\left(
{\bm \nabla_{\bf r}} n_{F}(\xi_{m}({\bf r})) \cdot {\bm \nabla_{\bf r}}
\right)
{\bm \nabla_{\bf r}}V
\right]
\times \hat{{\bf z}}
\right\}
.
\label{j2local}
\end{eqnarray}
Finally, the terms originating from the contribution ${\bf j}^{(3)}$ yield the
following second-order correction to the current density

\begin{eqnarray}
{\bf j}^{(3)}({\bf r})
&=&
\frac{e}{h}
\frac{l_{B}^{2}}{2}
\sum_{m=0}^{+ \infty}
\hat{{\bf z}}
\times
\left\{
(m+1) n'_{F}(\xi_{m}({\bf r})) \left[ \Delta_{\bf r}
V {\bm \nabla_{\bf r}} V - {\bm \nabla_{\bf r}} \left( \left|{\bm \nabla_{\bf r}} V \right|^{2} \right)\right]
+
\left(
\frac{2}{\hbar \omega_{c}}
 n_{F}(\xi_{m}({\bf r}))
-n'_{F}(\xi_{m}({\bf r})) \right)
\Delta_{\bf r} V {\bm \nabla_{\bf r}} V
\nonumber \right.
\\ &&
\left.
+
\left(
\frac{ n'_{F}(\xi_{m}({\bf r}))}{\hbar \omega_{c}} -
\frac{ n''_{F}(\xi_{m}({\bf r}))}{2}
\right)
 \left|{\bm \nabla_{\bf r}} V \right|^{2} {\bm \nabla_{\bf r}} V
- (m+1) n_{F}(\xi_{m}({\bf r})) \Delta_{\bf r} {\bm \nabla_{\bf r}} V
+
\Delta_{\bf r}
\left[
n_{F}(\xi_{m}({\bf r})) {\bm \nabla}_{{\bf r}} V
\right]
\right\}
.
\label{j3local}
\end{eqnarray}
\end{widetext}

\end{document}